\documentclass{elsart4-1}

\usepackage{graphicx}
\usepackage{epsfig}
\graphicspath{{pictures/},{.}}

\usepackage{amssymb}
\usepackage{color}

\usepackage[english,francais]{babel}

\newtheorem{e-proposition}[theorem]{Proposition}

\newtheorem{e-definition}[theorem]{Definition\rm}

\renewcommand{\theequation}{\arabic{equation}}
\usepackage{ulem}

\def\og{\leavevmode\raise.3ex\hbox{$\scriptscriptstyle\langle\!\langle$~}}
\def\fg{\leavevmode\raise.3ex\hbox{~$\!\scriptscriptstyle\,\rangle\!\rangle$}}

\begin{document}

\centerline{Astrophysics}
\begin{frontmatter}



\selectlanguage{english}
\title{Gamma rays as probes of the Universe}


\selectlanguage{english}
\author[a]{Dieter Horns}
\author[b]{\and Agnieszka Jacholkowska}
\ead{dieter.horns@physik.uni-hamburg.de}
\ead{Agnieszka.Jacholkowska@lpnhe.in2p3.fr}
\address[a]{Universit\"at Hamburg, Institut f\"ur Experimentalphysik,Luruper Chaussee 149\\ D-22761 Hamburg, Germany}
\address[b]{LPNHE, Universit\'e Pierre et Marie Curie Paris 6, Universit\'e Denis Diderot Paris 7, CNRS/IN2P3, 4 Place Jussieu\\
75252 Paris CEDEX 05, France}

\medskip
\begin{center}
{\small Received *****; accepted after revision +++++}
\end{center}

\begin{abstract}
The propagation of $\gamma$ rays over very large distances provides new insights on the intergalactic medium
and on fundamental physics. On their path to the Earth, $\gamma$ rays can annihilate with
diffuse infrared or optical photons of the intergalactic medium, producing $e^+ \, e^-$ pairs. The density of
these photons is poorly determined by direct measurements due to significant
galactic foregrounds. Studying the absorption of $\gamma$ rays from extragalactic
sources at different distances allows the density of low-energy diffuse photons to be measured.
Gamma-ray propagation may also be affected by new phenomena predicted by extensions of
the Standard Model of particle physics. Lorentz Invariance is violated in some models
of Quantum Gravity, leading to an energy-dependent speed of light in vacuum. From
differential time-of-flight measurements of the most distant $\gamma$-ray bursts and of flaring active
galactic nuclei, lower bounds have been set on the energy scale of Quantum Gravity.
Another effect that may alter
$\gamma$-ray propagation is predicted by some models of
String Theory, namely the mixing of the $\gamma$ ray with a light fundamental boson called
an ``axion-like particle'', which does not interact with low-energy photons. Such a mixing would
make the Universe more transparent to $\gamma$ rays than what would otherwise be,
in a sense it decreases the amount of modification to the spectrum that comes from the extragalactic background light.
The present status of the search for all these phenomena in
$\gamma$-ray astronomy is reviewed.\\
{\it To cite this article: Horns, D. and Jacholkowska, A., C.R. Physique XX (2016).}

\selectlanguage{francais}
\noindent{\bf R\'esum\'e}\\
\noindent
La propagation des photons $\gamma$ sur de tr\`es grandes distances nous permet de sonder le milieu intergalactique
et fournit des tests de physique fondamentale. Au cours de leur chemin vers la Terre,
ceux-ci peuvent s'annihiler avec les photons infrarouges et optiques du milieu intergalactique,
produisant ainsi des paires $e^+ \, e^-$. L'absorption des photons $\gamma$ \'emis par des sources extragalactiques \`a
diff\'erentes distances permet de mesurer ce fond diffus, par ailleurs tr\`es mal connu par des mesures directes en raison
des importants rayonnements d'avant-plan dus \`a la Galaxie. La
propagation des photons $\gamma$ peut aussi \^etre affect\'ee par de nouveaux ph\'enom\`enes
pr\'edits par des extensions du mod\`ele standard de la physique des particules. L'invariance
de Lorentz est viol\'ee dans certains mod\`eles de gravit\'e quantique o\`u
la vitesse de la lumi\`ere dans le vide varie avec l'\'energie du photon $\gamma$. Les
mesures diff\'erentielles de temps de vol sur les sursauts $\gamma$ et sur
les \'eruptions de noyaux actifs de galaxie ont permis d'obtenir des bornes
inf\'erieures sur l'\'echelle d'\'energie de la gravit\'e quantique.
Un autre effet pouvant affecter la propagation des photons $\gamma$ est pr\'edit par
des mod\`eles de la th\'eorie des cordes. Il s'agit du m\'elange quantique
entre le photon et une particule l\'eg\`ere de type ``axion'' qui n'interagit pas avec
les photons infrarouges et optiques. En diminuant dans les spectres l'empreinte de l'absorption
par le fond diffus, ce m\'elange rendrait l'Univers plus transparent que pr\'evu
aux photons $\gamma$.
L'article pr\'esente l'\'etat actuel des recherches sur l'ensemble de ces ph\'enom\`enes en astronomie $\gamma$. \\
{\bf {\it Pour citer cet article~: Horns, D. et Jacholkowska,
A.,  C. R. Physique XX (2016).}}
\vskip 0.5\baselineskip
\noindent{\small{\it Keywords~:} gamma rays; extragalactic background light; Lorentz invariance violation; axion-like particles}
\vskip 0.5\baselineskip
\noindent{\small{\it Mots-cl\'es~:} rayons gamma; fond extragalactique infrarouge et optique; violation de l'invariance de Lorentz;
axions}
\end{abstract}
\end{frontmatter}


\selectlanguage{english}

\section{Introduction}
\label{PROBES:intro}

Gamma rays are powerful and important messengers from cosmic accelerators.
In addition, the propagation of $\gamma$ rays over very large distances offers
a number of opportunities to investigate the extragalactic medium and to discover or
constrain new phenomena predicted by some extensions of the Standard Model of particle physics.
In this review, we focus on the following aspects:
\begin{itemize}
\item diffuse photons in the intergalactic medium with which $\gamma$ rays may annihilate
producing $e^+ e^-$ pairs, including the well-known cosmic microwave background (CMB), which
affects only the propagation of ultra-high energy $\gamma$ rays (PeV range), and
the diffuse background of optical and infrared photons --- the extragalactic background light (EBL) ---
which is poorly known from direct measurements, due to important galactic foregrounds;
\item the effects of a possible Lorentz-invariance violation (LIV), as predicted by models of Quantum Gravity (QG);
\item the mixing of photons with light
fundamental bosons ($m~c^2=\mathcal{O}(\mathrm{neV})$) such as axion-like particles (ALPs), as predicted by models of String Theory,
an effect sensitive to the magnetic field strength in extragalactic space.
\end{itemize}

Gamma rays propagating over very large distances are particularly good probes of the last two effects.
Lorentz-invariance violation, leading to an
energy-dependent dispersion of photons, would affect
propagating photons with a large product of energy $E$ and distance $d$.
Gamma-ray sources at cosmological distances reach values of $E\cdot d \approx
\mathrm{TeV~Gpc}$, challenging high-precision measurements in the laboratory.
The mixing of photons with ALPs depends on the
product of the magnetic field strength $B$ and the propagation distance as shown in
section~\ref{PROBES:sub:pheno_ALPS}. The largest values of the product $B \cdot d$
provide the highest sensitivity to this effect.
Even with extremely low intergalactic magnetic fields ($\sim 10^{-13}$~T)
this product can reach $B\cdot d \approx 3\times 10^{11}~\mathrm{T~m}$
with very distant sources,
a value which is currently not achievable in laboratory environments.

While the production of $\gamma$ rays from self-annihilating
massive dark matter probes fundamental
physics at the mass scale of supersymmetric particles, i.e.
close to the electro-weak symmetry-breaking scale \cite{cosmo}, searches for axion-like
particles are sensitive to new physics at intermediate energies of $10^{11}$ GeV and
processes leading to Lorentz invariance violation are expected to be related
to the Planck scale ($E_{Pl}=\sqrt{\hbar c^5/G}\approx 10^{19}$~GeV). Therefore, the
propagation of $\gamma$ rays from distant sources brings new insights
on presently unexplored energy domains of particle physics.

In section \ref{PROBES:sources}, those $\gamma$-ray sources most commonly
used in the search for LIV and for photon-ALP mixing are reviewed.
Section \ref{PROBES:propagation} describes $\gamma$-ray propagation
in the presence of extragalactic background light and of
magnetic fields. The phenomenology of LIV and the relevant
constraints from $\gamma$-ray observations are presented in section
\ref{PROBES:sub:pheno_LIV_QG}, whereas the current
understanding of photon-ALP mixing (PAM) and corresponding constraints
are discussed in section \ref{PROBES:sub:pheno_ALPS}.

\section{Sources of high-energy photons used in the study of propagation effects}
\label{PROBES:sources}

High-energy (HE, $E>100~\mathrm{MeV}$)
and very-high-energy (VHE, $E>100~\mathrm{GeV}$) photons are
produced in the framework of the most violent events occurring in the
Universe such as stellar
explosions or accretion on supermassive
black holes.
The relevant acceleration and radiation processes in the three types of sources
considered here are briefly described below:

\begin{enumerate}
\item
{\bf Gamma-ray bursts (GRBs)}: bright and short flashes of $\gamma$-rays are produced following the
collapse of a massive star or the coalescence of compact objects in a binary system.
This catastrophic event results in a strongly collimated relativistic plasma outflow in which multiple shells travel at different velocities,
producing internal shocks. Charged particles accelerated in these shocks further radiate HE photons \cite{grb}.
Due to the relativistic bulk flow of the emitting plasma,
photons are strongly beamed in the reference frame of the observer
who also experiences an apparent rapid variability. The prompt
emission of photons lasting from less than a second up to a few minutes is of
interest for the studies of fundamental physics. It is characterized by high
fluxes in the MeV-GeV energy range, with a variability in time of the
order of tens of milliseconds. The prompt emission is followed by counterparts
at longer wavelengths covering longer periods (up to months or years).
So far, GRB emission has not been detected at VHE energies.

\item
{\bf Active galactic nuclei (AGN)}: the accretion of matter onto
supermassive
black holes at the centre of some galaxies is at the origin of
their activity. HE and VHE $\gamma$ rays
originate from a strongly beamed emission in a jet of highly relativistic
plasma, most probably via inverse Compton scattering. AGN are variable sources
which are observed in two different regimes: steady state (i.e. slow evolution),
and flaring states in which strong variations of the $\gamma$-ray flux are observed
on periods from minutes to months \cite{agn}. Gamma-ray energy spectra from these objects suffer
an exponential cut-off at the high-energy end, in the TeV regime. This cut-off is mostly related to
the absorption of VHE photons through pair-production as discussed in section~\ref{PROBES:sub:radiation_transport}.
Only AGN with their jet pointing towards
the Earth (Blazars) can be used efficiently in LIV and PAM studies.
\item
{\bf Galactic Pulsars}:
HE photons are produced around magnetized rotating neutron
stars, remnants of ssupernovae explosions. They are
emitted by electrons accelerated to ultra-relativistic energies (Lorentz factors
of $10^7$) in the magnetosphere of the rotating neutron star.
Several models have been proposed to describe the underlying processes, for a review see \cite{pulsars}. Although not very
distant, pulsars complement the studies with GRBs and AGN with their very regular
pulsed emission at the millisecond scale.
\end{enumerate}

Detailed descriptions of the emission and the acceleration processes for the
three types of sources can be found in both volumes of the present thematic issue on $\gamma$-ray astronomy \cite{pulsars,grb,agn}.

\section{Canonical propagation of high energy $\gamma$ rays}
\label{PROBES:propagation}
\subsection{Radiation transport of energetic photons in the Universe}
\label{PROBES:sub:radiation_transport}
The propagation of energetic photons at energies considered here ($10^{9}$~eV to
$10^{14}$~eV) is in most cases well approximated by the propagation in a classical vacuum.
However, at cosmological distances, noticeable effects
which are intimately linked to the properties of the intergalactic medium
(radiation and thermal background plasma) as well as to
the quantum electrodynamics (QED) vacuum, can be probed.  {Before considering modifications of the propagation by
effects related to physics beyond the Standard Model, we briefly summarize the
evolution of a monochromatic and linearly polarized beam propagating in the intergalactic medium.}\\

 The intergalactic medium at redshift $z$ is filled with a very dilute
and magnetized plasma
(electron number density $n_e \approx 0.24 ~\mathrm{m^{-3}} (\Omega_b/0.04) (z+1)^3$, with $\Omega_b =
\rho_b/\rho_c$ the average mass density of baryons in units of critical density
$\rho_c=3H^2/(8\pi G)$ for a Hubble constant $H=74 \, \mathrm{km~s^{-1}Mpc^{-1}}$). The plasma frequency
in the medium $\omega_\mathrm{pl}=(n_e e^2 m_e^{-1} \epsilon_0^{-1})^{1/2}$ characterizes wave-like excitations
of the fluid.

Besides the plasma, an optical/infrared
(extragalactic background light or EBL) and cosmic microwave (CMB) photon field with
number density
$n_\gamma=(n_\mathrm{CMB}+n_\mathrm{EBL})(z+1)^3$, with $n_\mathrm{CMB}\approx 4.1\times 10^{8}~\mathrm{m^{-3}}$
and $n_\mathrm{EBL}\approx 20~\mathrm{m^{-3}}$
is present throughout the
universe\footnote{We neglect for now
evolutionary effects for the EBL which are not of importance for the discussion but
are included in the calculation presented}.
Additional localized photon fields in the
vicinity of galaxies and galaxy clusters can be neglected from the radiation transport in most
circumstances \cite{Maurer12,2015MNRAS.446.2267F}.
The photon state propagating along the direction $x_3$ characterized by the vector of state amplitudes along the direction of the transverse field strength $(\epsilon_1, \epsilon_2)$    with
energy $\hbar \omega$ is described by the following equation (see e.g. \cite{raffelt88,horns2012aa}):
\begin{eqnarray}
 \left(-i \frac{d}{dx_3} + \omega + \mathcal{M}\right)
\left( \begin{array}{c}
\epsilon_1\\
\epsilon_2\\
\end{array} \right) &=& 0,
\end{eqnarray}
where, in the simplified case of a homogeneous magnetic field  with a transverse component $\mathbf{B}_T$,
the matrix
\begin{eqnarray}
\label{propa}
 \mathcal{M} &=& \left( \begin{array}{cc}
 \Delta_{11} & \Delta_{12}  \\
 \Delta_{21} & \Delta_{22}
\end{array}\right)
\end{eqnarray}
contains off-diagonal elements, which mix the two polarization states due
to Faraday rotation and vacuum birefringence \cite{heisenbergeuler}. 
For the energies considered here, we safely neglect
the Faraday rotation and only consider the effect of vacuum birefringence related to\footnote{For easier reading, natural
units, i.e. $\hbar=c=1$, are used in equation \ref{delqed}, and those immediately following.}:
\begin{eqnarray}
\label{delqed}
\Delta_\mathrm{QED}&=\omega \frac{\alpha}{45\pi} \, \left(\frac{\mathrm{B}_T}{B_c}\right)^2,
\end{eqnarray}
with $\alpha=e^2/(4\pi)$ the fine structure constant and $B_c=m^2/e\approx 4.4\times 10^{13}~\mathrm{G}$
the critical field.
The off-diagonal terms are related to $\Delta_\mathrm{QED}$ by:
\[ \Delta_{12}=\Delta_{21}=\frac{3}{2} \, \Delta_\mathrm{QED}\sin\varphi\cos\varphi \: \: \: \mbox{with}  \: \: \:
\cos\varphi=\mathbf{B}_T\cdot\mathbf{e}_1/B_T.\]
The diagonal terms include the effect of the background plasma
$\Delta_\mathrm{pl}=-\omega_\mathrm{pl}^2/(2\omega)$,
and of vacuum polarization:
\[ \Delta_{11}  =  \Delta_\mathrm{pl} + \left(\frac{7}{2} \, \cos^2\varphi + 2 \sin^2\varphi\right) \, \Delta_\mathrm{QED} \: \: \: \mbox{and}  \: \: \:
 \Delta_{22}  =  \Delta_\mathrm{pl} + \left(2\cos^2\varphi + \frac{7}{2} \, \sin^2\varphi\right)\, \Delta_\mathrm{QED}. \]
 Note that $\Delta_\mathrm{QED}\propto \omega B^2$ dominates over $\Delta_\mathrm{pl}$ for sufficiently energetic photons or
strong magnetic fields.

So far, the treatment of $\gamma$-ray propagation has not included inelastic interactions, like pair-production
processes with low-energy background photons ($\gamma + \gamma \rightarrow e^+ + e^-$).
{Given the low matter density, }
$\gamma$-ray interactions in the medium are dominated
by the latter process, characterized by the mean free path $\lambda_{\gamma\gamma}$, and additional interactions with
the background plasma are negligible.
Integrating the inverse of the mean free path of $\gamma$ rays
over the line of sight yields
the optical depth $\tau_{\gamma\gamma}=\int  \lambda_{\gamma\gamma}^{-1}(E) \, ds$; the initial number of photons
produced is thus reduced\footnote{Note that the integration over the path length $ds$ to obtain $\tau_{\gamma\gamma}$ is sensitive to the actual cosmological model
(see e.g., \cite{DomPrada}).} at the detector by a factor $\exp(-\tau_{\gamma\gamma})$.
In the standard scenario for photon propagation, $\lambda_{\gamma\gamma}^{-1}(E)$ is given by \cite{nikishov,1967PhRv..155.1404G}:
\begin{eqnarray}
\label{lambda}
\lambda^{-1}_{\gamma\gamma}(E) &=&
\int  \frac{dn_\gamma(\epsilon)}{d\epsilon} \, d\epsilon \int  \sigma_{\gamma\gamma}(\mu, \epsilon, E)\frac{1-\mu}{2} \, d\mu .
\end{eqnarray}
The first integration is over $\epsilon$, the energy of the background photon, and the second over
$\mu=\cos\theta$, $\theta$ being the angle between the propagation directions of the two photons,
assuming that the differential number density $dn_\gamma/d\epsilon$ of background photons is
isotropic. {The mean free path is $\mathcal{O}(100~\mathrm{Mpc})$ for energies above 10~TeV and quickly grows to $\mathcal{O}(\mathrm{Gpc})$
for energies $\approx 1$~TeV. At energies below 100 GeV, the universe is practically
transparent to $\gamma$ rays.}
The cross section $\sigma_{\gamma\gamma}$ averaged over the angle $\theta$ peaks at a value of $\epsilon E/m_e^2 \approx 1.3$
which corresponds to a photon wavelength of $1.24~\mu\mathrm{m}/(E/ 1 \,\mathrm{TeV})$.
The differential number density $dn_\gamma/d\epsilon$ is well known for the
CMB and closely follows a thermal black-body spectrum, while it is rather difficult to measure or predict for
the EBL. According to the current understanding, a strict lower limit can be derived from counting faint sources with e.g. the
Hubble space telescope,
while a strict upper limit can be determined from measuring the large scale brightness of the sky, subsequently corrected by
subtracting the dominant foreground emissions, those of the Galaxy and the interplanetary medium.
For an extensive summary of the background photon field, see \cite{dwek_krennrich} and references therein.
An intermediate value, most likely close to the reality,
has been determined from the measurement of $\tau_{\gamma\gamma}(E)$ using $\gamma$-ray spectra measured with the Fermi-LAT instrument \cite{ackermann2012} and with
the H.E.S.S. Cherenkov telescopes \cite{abramowski2013} (see Section~\ref{ebl-constraint}).

To account for this absorption, the propagation matrix of equation~\ref{propa} is modified by adding  $-i\lambda^{-1}_{\gamma\gamma}/2$ to the diagonal terms.
The standard picture of pair-production leading to absorption is however a simplification, as
the pairs produced are subsequently cooling either radiatively via synchrotron
or inverse Compton effects or non-radiatively via plasma instabilities induced
by the pairs propagating in the background plasma \cite{Broderick2012}.
While it is not yet entirely clear which effect(s) dominate(s), there are two limiting cases:
\begin{enumerate}
 \item \textit{Pair production cascade:} The pairs produce energetic photons via inverse Compton upscattering of the background radiation field (mainly CMB).
These photons subsequently continue to increase the number of particles at the expense of reducing the average energy per particle:
a  pair/inverse-Compton cascade develops and channels the energy of the primary $\gamma$ ray into a large number of lower-energy photons with energies below 100 GeV
for which $\lambda_{\gamma\gamma}(E)$ exceeds a few Gpc.
 \item \textit{Quenched cascade:} In this case, the cascade is suppressed by pairs cooling predominantly through synchrotron
radiation rather than by inverse Compton scattering ($\dot E_\mathrm{syn}/\dot
E_\mathrm{IC}=u_B/(u_\mathrm{CMB}+u_\mathrm{EBL})>1$).
However, when comparing the magnetic energy density:
\[ u_B=\frac{B^2}{2\mu_0}=0.024~\mathrm{eV~m^{-3}} \left(\frac{B}{10^{-13}~\mathrm{T}}\right)^2  \: \: \: \mbox{with that of the CMB}  \: \: \:
u_\mathrm{CMB}=2.5\times10^{5}~\mathrm{eV~m^{-3}} (z+1)^{4},\]
this is  obviously
only of concern in regions with increased magnetic field strength (clusters of galaxies or environments of active galactic nuclei). \\
For powerful $\gamma$-ray sources with a $\gamma$-ray luminosity $L_\gamma>10^{34}~\mathrm{W}$, the excitation of two-stream
instabilities by pairs propagating in the background
plasma can dominate the energy losses. The growth rate of the instabilities
$\propto \gamma L_\gamma/n_e^{1/2}$,
with $\gamma$ the Lorentz factor of the pairs, can lead to a suppression of cascade formation \cite{Broderick2012}.
\end{enumerate}
While energy losses through synchrotron emission (dominant in case (ii) above) are generally not
important, the subtle effect on the secondary $\gamma$-ray emission of the magnetic field deflecting the pairs
provides an observational test
of the intergalactic magnetic field by measuring the absorbed primary
$\gamma$-ray component as well as the lower-energy secondary component.
This in turn has been used to estimate lower bounds on the
magnetic field \cite{neronov2010}\footnote{For a review on the magnetic field in the
intergalactic medium, see \cite{durrer2013}}. However, non-radiative
cooling effects may weaken these limits. \\
The secondary effects of photon pair-production effectively modify the
dispersion relation, implying a change of the optical depth.
In some specific scenarios, the generated secondary photons may even
dominate the observed $\gamma$-ray spectra of distant sources \cite{essey2010}.

{Introduction of LIV into dispersion relations would lead to a modification of its real component
for sufficiently energetic photons.
On the other hand, both LIV and PAM act on absorption processes and modify the optical depth $\tau$ as discussed in subsections \ref{ebl-constraint}, \ref{PROBES:sub:pheno_LIV_QG}, and \ref{PROBES:sub:pheno_ALPS} respectively.
In consequence, LIV as well as PAM would lead in the most general case to a directional as well as distance dependence of
the dispersion relation.}
\subsection{Inferring the optical/near-infrared background light from $\gamma$-ray
spectra}
\label{ebl-constraint}
\begin{figure}
 \centering
 \includegraphics[width=14cm]{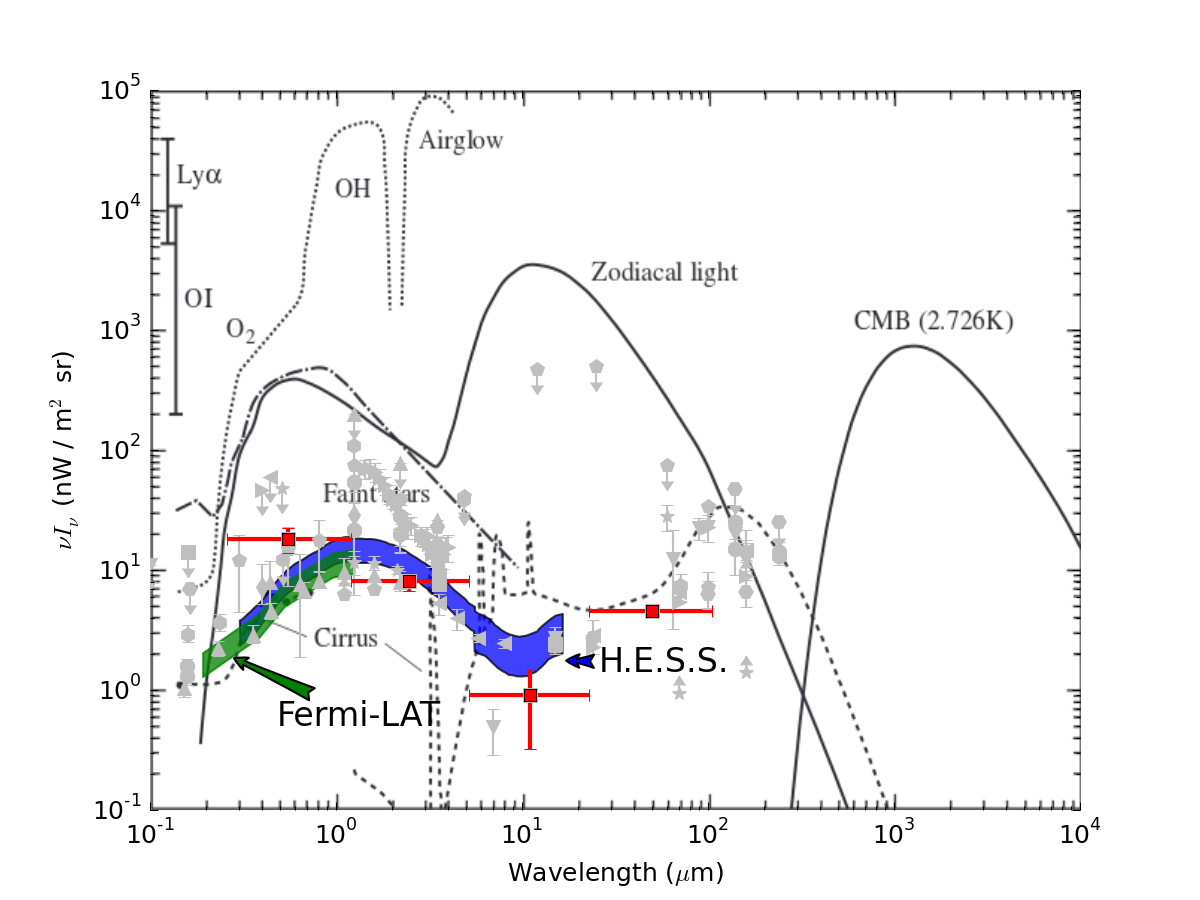}
 \caption{ The sky brightness from ultra-violet to cm-wavelengths (adapted from
\label{ebl}
\cite{leinert1998}). {To be noted} the large foreground emission in the optical to mid-infrared (at $\approx 10~\mu\mathrm{m}$) from the inter-planetary medium (zodiacal) as well as from the atmosphere. The contribution of direct
stellar light and stellar light scattered by
the interstellar medium (cirrus) as well as atmospheric glow are indicated as well. The measurements of the extragalactic background light (in grey and colored markers,
for a list of references used here, see the text and \cite{2012MNRAS.426.1097R})
falls in three categories: (i) estimates  from source counts
and fluctuation measurements, (ii) estimates from integral measurements,
and (iii), marked as green and blue bands and red markers with horizontal bars, estimates derived from absorption features
in $\gamma$-ray spectra. The estimates (i) and (ii) have to be considered as lower
and upper limits respectively, given that the source counts are not complete
and would miss a truly isotropic component and (ii) suffer from obvious large
foregrounds and from unresolved source contributions.}
\end{figure}

In intergalactic space, the integrated light
(ultra-violet, optical, infrared) emitted by
distant sources (first stars, galaxies in the early universe, and
any potentially unknown radiation process),
forms a mostly isotropic background glow.  This radiation component
is not to be confused with the cosmic microwave background which is of
primordial origin. It carries however vital information on the star formation history of the Universe as well as on the amount of ``shining'' matter.
Any, at this point unknown, light-producing mechanism (first stars, decaying
particles) which may have been important in the early universe could have
left its imprint in the spectral energy distribution of this light
\cite{2004PhR...402..267O,2009A&A...498...25R,2012ApJ...745..166M,2012MNRAS.426.1097R,2012MNRAS.420..800G}.
 For an observer on Earth, this extragalactic background light (EBL) is
difficult to measure directly as it is small with respect to more
prominent foreground emission from the planetary and inter-stellar medium
(see Fig.~\ref{ebl} for the spectral energy distribution of the night-sky brightness).
The most common approaches have been either related to {the}
counting of faint sources up to
the instrumental sensitivity limit\footnote{Unresolved sources contribute to
the noise spectrum}
(e.g., \cite{1996ApJ...470..681K,2000ApJ...528...74K,2000MNRAS.312L...9M,2000A&A...355...17L,2002A&A...384..848E,2003A&A...407..791M,2004ApJS..154...87D,2004ApJS..154...39F,2004ApJS..154...70P,2006A&A...451..417D,2006AJ....131..250F,2010A&A...512A..78B,2010A&A...518L..30B,2011ApJ...736...80V,2011ApJ...737....2M,2013A&A...553A.132M}) or
measuring the integrated emission subsequently corrected by the model-dependent subtraction of the
foreground emission
(e.g., \cite{1990IAUS..139..257M,1998ApJ...508L...9D,1998ApJ...508...25H,2000ApJ...544...81F,2000ApJ...536..550G,2000ApJ...545...43W,2005ApJ...626...31M,2000AJ....120.1153B,2002ApJ...571...56B,2005ApJ...632..713B})\footnote{For further references, see \cite{dwek_krennrich}}.
Both approaches suffer from
the inherent problem that they should be interpreted
either {as} a lower-limit (given the
limited instrumental sensitivity) or as an  upper limit (given that the
measurement is {done in an integrated way} and the foreground emission dominates). Claims
for detection have been controversial and not consistent with each other,
highlighting the systematic uncertainties {which are} present. \\

A promising indirect method which is not contaminated by the
disturbing foreground emission is related to the close link between the photon
number density in the intergalactic space and the photon-pair absorption (see
equation~\ref{lambda}).
Recalling that the
peak in the pair-production rate is reached for a wavelength of $1.24~\mu\mathrm{m}/(E/1 \, \mathrm{TeV})$,
it is clear that the EBL-induced absorption mostly affects observations in the energy band from 100 GeV to 100 TeV.
With the advent of $\gamma$-ray measurements from the
ground, observations of $\gamma$ rays
up to TeV energies are possible, where absorption through pair-production processes
should start to affect the energy spectra.\\
The discovery of $\gamma$-ray emission
above 10 TeV from
low redshift blazars (Mkn~421 and Mkn~501  at $z=0.03$)
were used to derive the first meaningful constraint on the level of the EBL
in the mid-infrared (e.g., \cite{1993ApJ...415L..71S,1994Natur.369..294D}). \\
The problem of determining the EBL spectrum $n_\gamma$ by estimating the optical depth $\tau_{\gamma \gamma}(E)$ from the measured spectrum
requires however some assumptions.
The early efforts to derive constraints (upper limit on $\tau_{\gamma \gamma}(E)$) were
troubled by rather strict assumptions on the shape of the intrinsic spectrum
of the $\gamma$-ray sources
as well as on the spectrum of the background light $n_\gamma$.
The most common procedure used is to vary the level of the EBL until the
corrected $\gamma$-ray spectrum shows an apparent unphysical behavior
\cite{1995ApJ...445..227B,1998APh.....9...97F,2001A&A...371..771R,2005ApJ...628..617S,2006Natur.440.1018A}. \\
More elaborate approaches include the use of multiple source spectra
\cite{2005ApJ...618..657D,2011ApJ...733...77O,2014ApJ...795...91S} or
multi-wavelength observations and models of the $\gamma$-ray spectra to reduce the
uncertainties related to the unknown intrinsic spectrum \cite{2000A&A...359..419G,2013ApJ...770...77D,2014ApJ...788..158A}. \\
Additionally, the shape of the
EBL spectrum has been
left to vary freely in order to eliminate a model uncertainty
\cite{2009ApJ...698.1761F,2007A&A...471..439M}, and the energy range has been extended
by including Fermi-LAT observations \cite{2010ApJ...714L.157G,2010A&A...522A..12Y,2012A&A...542A..59M}.
The result of these approaches have been
upper limits on the intensity of the EBL
which exclude some integrated measurements and start to close-in on the
lowest possible level of the EBL set by optical source count measurements
with the Hubble space telescope.
 \\
In recent years, the $\gamma$-ray energy spectra of AGN have
considerably improved in number and in precision. The approaches described above
are based upon a comparison of a reconstructed intrinsic spectrum with some
assumption on the allowed shape at the emission. This in
turn has been used to constrain the maximum optical depth and therefore the level of $n_\gamma$
under some assumption of its shape. With the improved energy spectra, it
has become feasible to search for features related to
absorption, e.g. a gradual softening of the energy spectrum. Since this
effect is related to the distance of the source, source-intrinsic effects can be
ruled out as being responsible for the observation of an increased curvature
with increasing redshift of the source.\\
By fixing the EBL spectrum shape
to a fiducial model-dependent one and leaving only a normalization factor $\zeta$ to vary,
($n_\gamma = \zeta \cdot n_\mathrm{fid}$), the characteristic
softening of the energy
spectrum depends only on the value of $\zeta$.  By combining various energy spectra from sources at different $z$, this method allows $\zeta$
to be measured and the intrinsic spectra to be reconstructed at the same time. \\
This method has been successfully applied to H.E.S.S. observations of AGN providing the first measurement of the EBL level \cite{abramowski2013}. The result, consistent with
previous upper bounds, is found about $20~\%$ above the lower limit from source counts (see the blue band in Fig.~\ref{ebl}).
In parallel to the EBL estimate from ground-based observations,  the
large sample of $\gamma$-ray emitting BL Lac objects detected with
the Fermi-LAT instrument has been used to determine a value of $\zeta$ in
a similar approach \cite{ackermann2012,2013ApJ...772L..12G}.  The two independent measurements
agree quite well (see Fig.~\ref{ebl})
and complement each other since the ranges of the covered wavelengths are partly overlapping.
\\
Results of a recent study deriving the EBL from $\gamma$-ray spectra of
38 sources are also shown in Fig.~\ref{ebl} by red markers
with horizontal bars \cite{biteauwilliams2015}. The resulting EBL estimate shows
notable deviations from model predictions (especially in the lowest band centered at 0.5~$\mu$m)
and is in tension with lower bounds from source counts in the
mid infrared. Future studies and observations will clarify these discrepancies
(see section  \ref{anomalous} on possible interpretation of this result in
the context of photon-ALP mixing).

{Measurements of the opacity of the Universe to the HE and VHE $\gamma$ rays emitted by extragalactic sources
and the knowledge
of the underlying processes contributing to the EBL effects are of great importance for standard astrophysics and
for physics beyond the Standard Model of particle physics and cosmology.
The studies of the EBL phenomena are necessary for AGN and GRB source modeling; they lead to the evaluation of
the diffuse radiation fields and of the magnetism of the Universe at large scales and affect
LIV and ALP search analyses, as presented in the following sections.}

\section{Modifications of the canonical $\gamma$-ray propagation}
\subsection{Phenomenology of Lorentz-invariance violation}
\label{PROBES:sub:pheno_LIV_QG}

Lorentz Invariance (LI) is a basic component of Einstein's Special Relativity.
It is strictly valid
in Quantum Mechanics and has been verified in various accelerator experiments at the
electro-weak scale. It is currently assumed that its applicability ranges
up to energies close to the Planck energy, $E_{Pl} \simeq 1.22\times10^{19}$ GeV,
above which known laws of physics are supposed
to break down. Testing LI with cosmic $\gamma$ rays allows probing the
full domain of its applicability in the highest observable energy regime.

On the other hand, Lorentz Invariance Violation (LIV) has also been largely
predicted in the framework of various classes of Quantum Gravity (QG) models.
This effect should occur at the Quantum Gravity energy scale $E_{QG}$, expected to be of the
order of the Planck scale~\cite{Planck}, (in
some cases lower, e.g. in some Loop Quantum Gravity
models~\cite{Solodukhin,Rovelli}, or in some string-theory (M-theory) models).


Tests of LIV with high-energy
photons from distant sources were first proposed by~\cite{Amelino2}~\cite{Ellis1},
according to a scenario of brane-type models~\cite{Ellis1} in which
QG effects result in an anomalous refractive index. During their
propagation through the extragalactic medium on cosmological distances, photons from distant
astrophysical sources, such as $\gamma$-ray bursts (GRBs) and active galaxies, may
cumulate tiny QG effects due to the foamy structure
of the quantum vacuum. As photon energies are spread over a large range, these
QG effects can be detected as corrections to the standard energy-momentum relation written here for real photons.
Equation~\ref{eq:disp} shows a simple modification of this relation
due to LIV terms:

\begin{eqnarray}
\label{eq:disp}
E^2\simeq p^2c^2 \times \left[ 1-\sum_n s_{\pm} \left(\frac{E}{E_{\rm QG}}\right)^n \right]
\end{eqnarray}
where $c$ is the usual speed of light (i.e. at the limit of zero photon energy),
$s_{\pm}$ is a theory-dependent factor equal to $+1$ $(-1)$ leading to
a decreasing (increasing) photon speed with increasing photon energy, corresponding to
the ``subluminal'' or ``superluminal'' case. For $E \ll E_{QG}$, the lowest order
term in the series not suppressed by the theory is expected to dominate the sum. If
the $n = 1$ term is suppressed, e.g. if some symmetry law
is involved, the next term $n = 2$ will dominate.

The power of a given source to constrain LIV increases with its distance, its variability
in time and the hardness of its energy spectrum. The first two qualities for a given source provide
the best limits on the LIV $n = 1$ parameter (case of the high-redshift GRBs), whereas the number
of VHE photons plays a crucial role in the determination of the $n = 2$ term (case of TeV blazars).
The next generation of the ground-based Atmospheric Cherenkov Telescopes (ACTs)~\cite{FUTURE}
will be able to take advantage of these three factors more efficiently.
The practical way to detect the correction terms in
formula \ref{eq:disp} consists of measuring the arrival time and energy of each photon,
which allows energy-dependent light curves of the source to be reconstructed, from which a search for a possible time lag
between them can be made.
For sources at cosmological distances, the analysis of time
lags as a function of the redshift requires a correction due to the expansion of
the universe~\cite{Jacob}, which depends on the cosmological model.
For a given value of the parameter $n$ in equation \ref{eq:disp} ($n=1$ or 2),
the mean value of arrival time delays $\Delta t$ between photons emitted at the
same time at the source and detected with energies $E$ and $E+\Delta E$ respectively, is related to
$\Delta E$ by the equation:
\begin{eqnarray}
\label{eq:taun}
\tau_n\equiv \frac{{\rm\Delta} t}{{\rm\Delta} E^n} \approx s_{\pm}\frac{(n+1)}{\mathrm2{E}_{\rm QG}^n \mathrm{H}_\mathrm{0}} \int_0^z \frac{(1+z')^n\,dz'}{\sqrt{\Omega_m(1+z')^3 + \Omega_{\Lambda}}}\end{eqnarray}
According to~\cite{Bahcall}, the cosmological
parameters were set to $\Omega_m = 0.3$, $\Omega_{\Lambda} = 0.7$ and
$\mathrm{H}_\mathrm{0} = 2.3\times10^{-18}$~s$^{-1}$ in most of the studies considered here.


 Various data sets provided by HE and VHE $\gamma$-ray astronomy have been used to
set limits on the energy scale at
which Quantum Gravity effects causing LIV may arise. Untill now, mainly
time-delay measurements as a function of photon energy
were performed assuming a deterministic time-of-flight vs. energy relation.
Stochastic (or ''fuzzy'') effects postulated by some of the QG models
(\cite{Ellis1} and references therein) have largely not been investigated yet. Therefore,
only deterministic time-dispersion results are discussed here and constraints are set
on LIV-induced time delays described by the linear or quadratic term in relation~\ref{eq:disp}.
These limits were translated into mainly one-sided
$95$\% confidence limits (CLs) on the QG energy scale in case of linear and quadratic scenarios.
The CL calculations followed usual procedures depending on the method used for
the extraction of results. In most of the earlier publications,
the contributions of systematic errors were not taken into account and the statistical treatment was not always rigorous.
Only the latest analyses of a giant flare of PKS 2155-304
and of four remarkable Fermi GRBs (GRB 080916C, GRB090510, GRB090902B, GRB090926A) took these aspects into account
using a semi-bayesian methodology, i.e. mixing frequentist and bayesian approaches in different steps.

Results on time-delay measurements and on the corresponding constraints on the
$E_{QG}$ during the last 10 years were obtained from
satellite experiments and ground-based Cherenkov telescopes
observing various sources, such as
AGN, GRBs, and pulsars. Indeed,
the potential of each time-delay measurement depends on the following conditions:
\begin{itemize}
\item the nature of the source, particularly, its variability time-scale;
\item the detector performance, i.e. acceptance, angular and energy resolutions;
\item the analysis method used to extract possible LIV effects, while controlling systematic errors due to the experimental setup as well as to the nature of the source.
\end{itemize}

\begin{table}
\caption{General properties of sources of interest in use for LIV studies.}
\begin{center}
\begin{tabular}{|l|c|c|c|}
\hline
Property & GRBs & AGN & Pulsars  \\
\hline
Redshift        & $ < 8.2$   &  $ < 0.8$ & ~0  \\
Energy range     & $ < 100$ GeV   &  $ < 10$ TeV & $ < 300$ GeV \\
Relevant time scales    & 10-100 ms  &  1-10 min  & 10 ms\\
Intrinsic effects    & known  &  moderate  & under control\\
Best results      & GRB 090510    &  PKS 2155-304  &  CRAB \\
    &  (Fermi)   &  (H.E.S.S.)  &  (VERITAS)\\
\hline
\end{tabular}
\end{center}
\label{tab:Table1}
\end{table}

The impact of each aspect is briefly discussed in the following.
\begin{enumerate}
\item
 Importance of the variety of sources (GRBs, AGN and pulsars). Considering
different types of sources is crucial for LIV studies. In LIV effects,
photon energies are affected according to the source redshift,
which is not necessarily the case if observed delays refer to the emission
time at the source.
Therefore, population studies at different redshifts
allow to discriminate between the two
interpretations. Moreover, the coverage of a large energy
range (from MeV to TeV) strengthens
the overall conclusions from
these analyses. The important properties of the considered sources are
summarized in Table~\ref{tab:Table1} for GRBs, AGN and pulsars.
In particular, close millisecond pulsars offer the opportunity of combining a high variability and
a high duty-cycle of observations, unlike GRBs and AGN.
\item Detector response and performance. The quality factors for LIV studies are a large acceptance,
an excellent time resolution for the observation of transient events and a good energy resolution.
The best cases considered
up to now correspond to a short GRB detected
by Fermi-LAT (GRB 091015) providing an important
sample of photons in the GeV range
and to a giant flare of the active
galaxy PKS~2155-304 as observed in 2006 with H.E.S.S., yielding
thousands of $\gamma$ rays in the TeV range.
\item Analyses and methods dedicated to the search for LIV effects.
Table~\ref{tab:Table2} lists the
variety of methods which were developed and used in LIV studies. The most
accurate results came from studies based on maximum likelihood
methods. However, due to the importance of the results for physics, a cross-check
is mandatory with methods probing different aspects of the light-curves.
Systematic effects should also be addressed in a complete way
and their values propagated when deriving the QG limits. Indeed, the precision
measurements of the time-lags require excellent statistical calibration
procedures and a good understanding of systematic uncertainties.
\end{enumerate}

\begin{table}
\caption{Most successful methods developed for LIV studies and used in different analyses. }
\begin{center}
\begin{tabular}{|l|c|c|}
\hline
Method  & Experiment  & Remarks  \\
\hline
Cross-correlation function (CCF)      & BATSE (GRBs), H.E.S.S. (AGN)   &  Low systematics   \\
Energy Cost function (ECF)    & MAGIC (AGN)   &  Good precision \\
Wavelet transform (CWT)   & BATSE  HETE-2  SWIFT (GRBs), H.E.S.S. (AGN)  &  Dependence on LC binning \\
Likelihood fit   & INTEGRAL Fermi (5GRBs), MAGIC H.E.S.S. (AGN)   &  Most precise with low statistics  \\
Sharpness maximization (SMM)    & Fermi (GRBs)   &  Dependence on source effects  \\
Pair View (PV)  & Fermi (GRBs)   &  Good performance with moderate  statistics \\
\hline
\end{tabular}
\end{center}
\label{tab:Table2}
\end{table}

 \subsubsection{Experimental results from the astrophysical observations of GRBs, AGN and Pulsars}
\label{sec:results}

In the pre-Fermi era, considerable efforts were already dedicated to GRB measurements
in the keV-MeV range, e.g. by Lamon et al.~\cite{Lamon} with INTEGRAL, by
Bolmont et al.~\cite{Bolmont1} using HETE-2 detections of 15 GRBs and by
Rodr\'iguez-Mart\'inez et al.~\cite{Rodriguez} using Swift and Konus-Wind
observations of GRB 051221A. At that time, Ellis et al.~\cite{Ellis2}
combined the results from HETE-2, BATSE and Swift GRBs deriving the
limit $E_{QG} > 2.1~10^{16}$ GeV at $95$\% CL, still three orders of
magnitude below $E_{Pl}$.

\begin{figure}[!t]
  \centering
  \includegraphics[width=7.5cm]{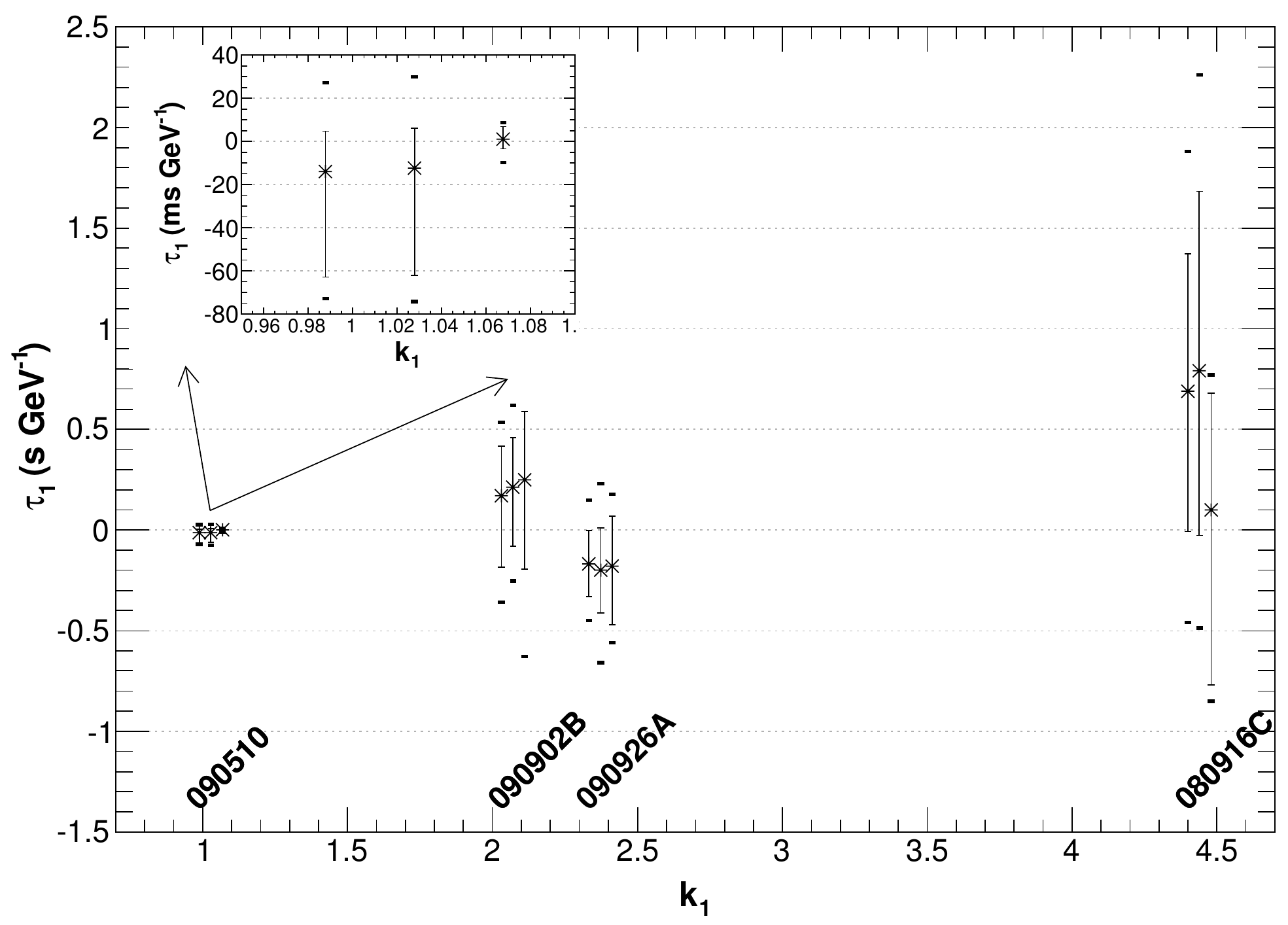}
  \includegraphics[width=7.5cm]{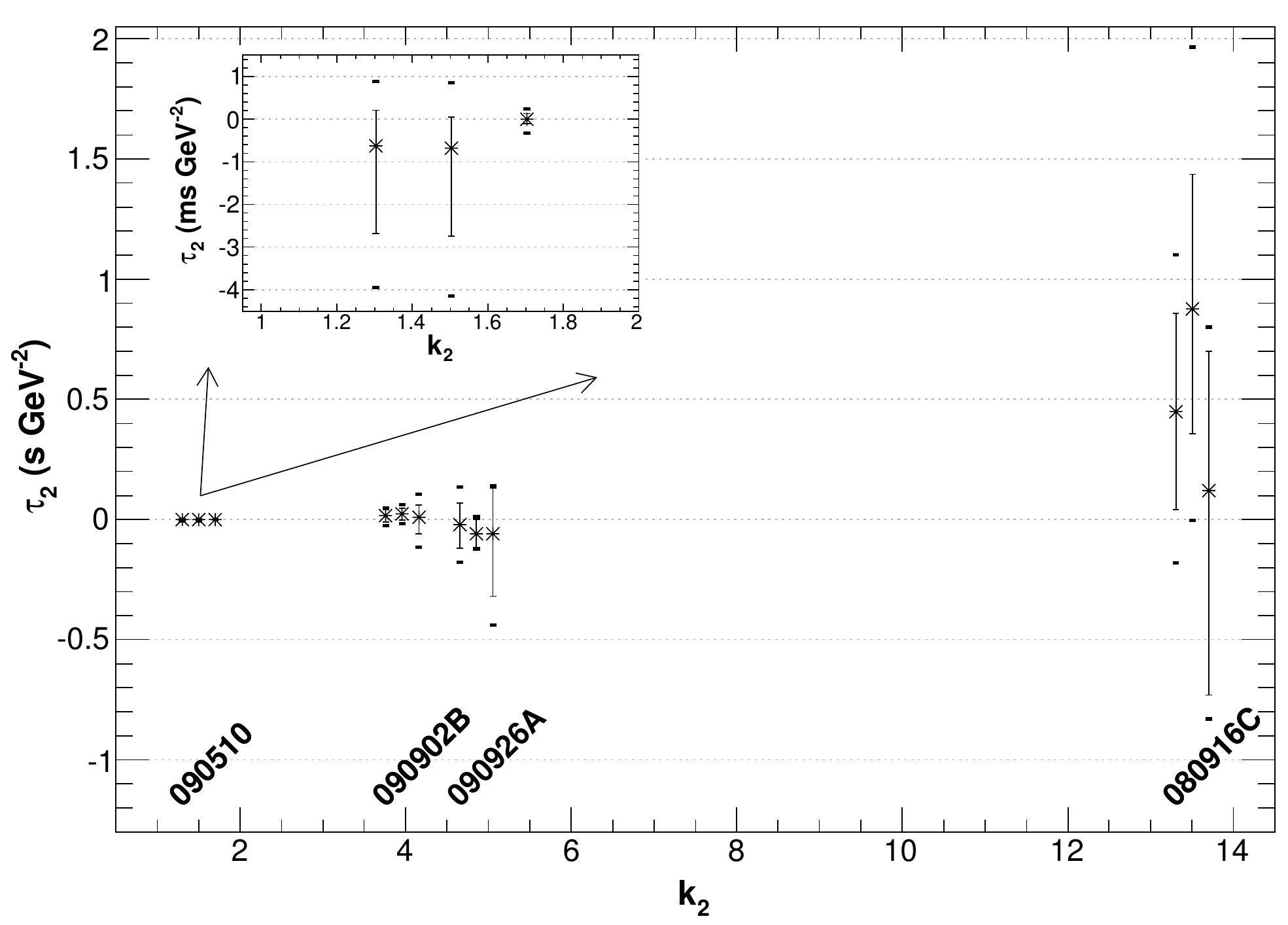}
  \caption{Results on the parameters $\tau_n$ defined in formula \ref{eq:taun} from four Fermi GRB, data when
applying three different methods in the analysis (left $n = 1$, right $n = 2$). The parameters in abscissa $k_1$ and $k_2$ are distance measures
taking account of the expansion of the Universe during photon propagation.
The most precise measurement was obtained with GRB 090510 with the likelihood approach. It should be noticed that the statistical errors increase with redshift of the source due to a more limited statistics of detected photons
  (from~\cite{Vasileiou}).}
  \label{fig:Figure1}
\end{figure}

The main break-through in time-of-flight LIV studies came with the
Fermi mission and the following detection of numerous GRBs with photon
energies up to a hundred GeV, taking advantage of
the unprecedented sensitivity of
the Fermi Large Area Telescope (LAT)~\cite{Piron,Atwood}. The
first publications providing constraints combine data from the Fermi-LAT and
those from the Gamma-ray Burst Monitor (GBM) on GRBs 080916C~\cite{Abdo1} and
090510~\cite{Abdo2} (see also Shao et al.~\cite{Shao} and Nemiroff et
al.~\cite{Nemiroff} using multiple Fermi GRBs). These data yielded limits on the
linear term in relation \ref{eq:disp} at the level of the Planck scale,
disfavoring several QG models which predict significant deviations at this level.
Following subsequent theoretical discussions, new studies were carried out on
four “golden” GRBs (GRB~080916C, GRB~090510, GRB~090902B, GRB~090926A), providing important samples of photons in the
GeV range, excellent variability in time, and redshift values distributed
between 0.9 and 4.3. More robustness on experimental
analyses was required at that time. This was fulfilled by cross-checking
the three most efficient statistical methods and a better control of
systematic uncertainties, resulting in a recent publication by Vasileiou et
al.~\cite{Vasileiou}. Fig.~\ref{fig:Figure1} shows the $\tau_{n}$ parameters of formula \ref{eq:taun}
($\tau_1$ for the linear case, $\tau_2$ for the quadratic term) obtained for the four GRBs with the three methods, each probing different
aspects of GRB light-curves. In Fig.~\ref{fig:Figure1}, the $k_n$ parameters ($n=1$ or 2) in abscissa are measures of the source distances which
take account of the expansion of the universe during the propagation of photons~\cite{Jacob}. Clearly, all values of $\tau_n$ are compatible with 0 for the four GRBs with the three methods\footnote{The error bars in the figure reflect internal properties of each GRB, in particular
the statistics of photons observed in the GeV range.
The most precise measurement comes from
GRB~090510 which presents the fastest
flux variations, at the level of few tens of millisecond. This measurement yields the most
stringent limit on the QG energy scale despite the low redshift value of
GRB~090510.}.

\begin{figure}[!t]
  \centering
  \includegraphics[width=7.5cm]{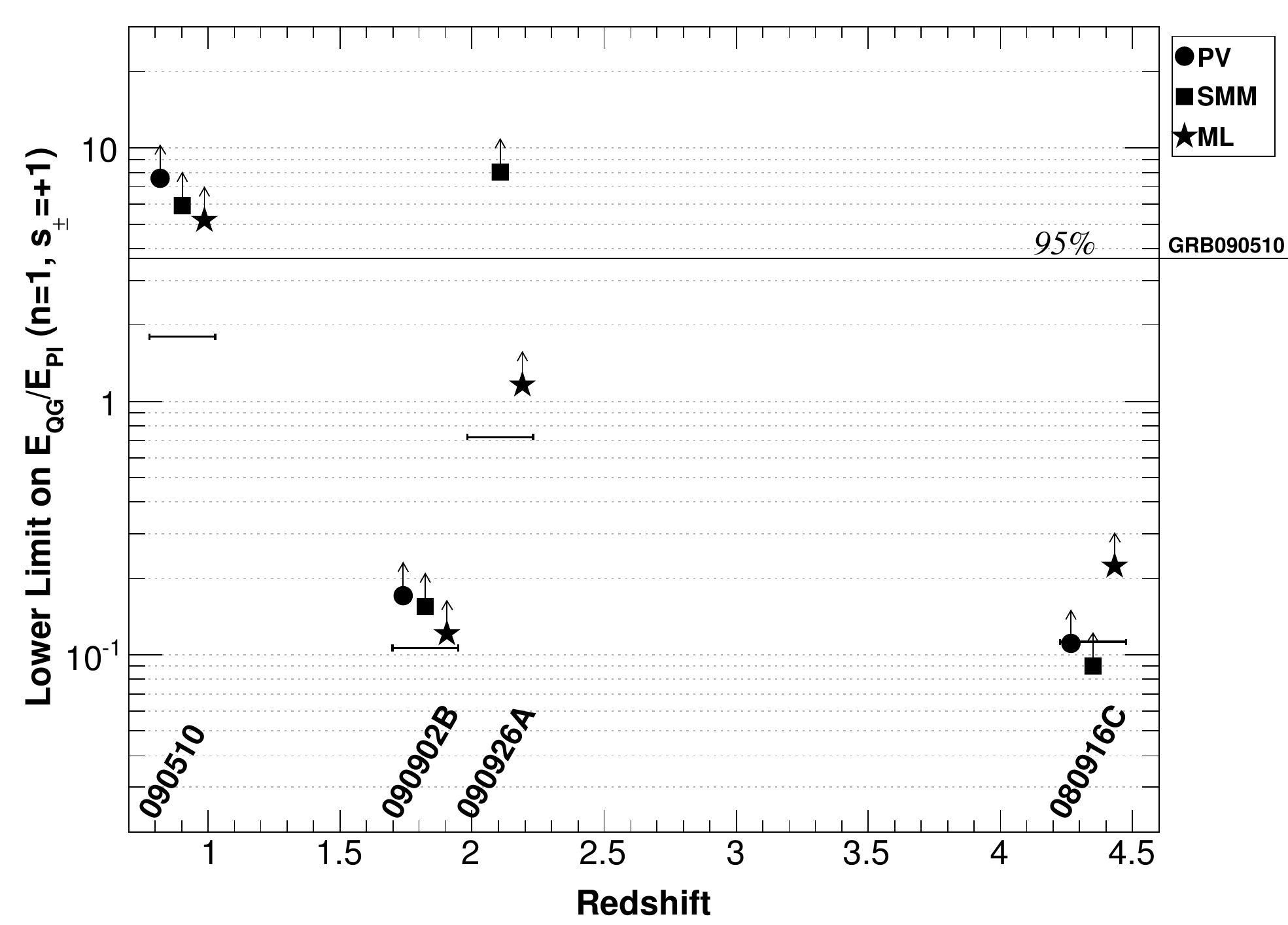}
  \includegraphics[width=7.5cm]{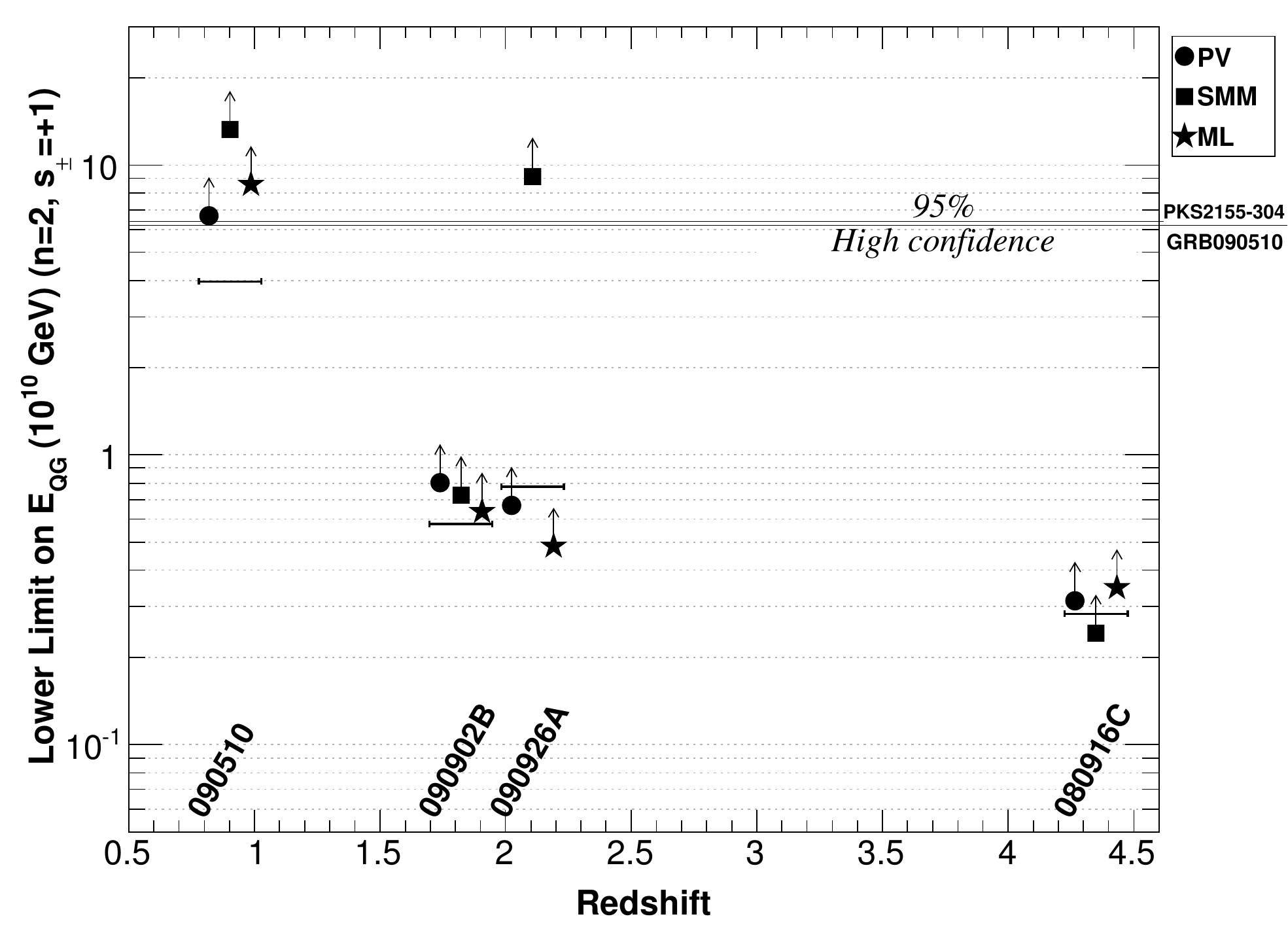}
  \caption{
 $95$\% CLs on QG energy scale with four Fermi GRB data when applying three different methods in the analysis (left n = 1, right n = 2). The most performant result was obtained with GRB 090510, well above  $E_{Pl}$. The horizontal bars for GRB reflects averaged over three method values of the CLs when the systematic errors have been included
  (from~\cite{Vasileiou}).}
  \label{fig:Figure2}
\end{figure}

Since no LIV effects have been detected in the preceding analysis, only lower bounds have been
set on the linear and quadratic terms of formula \ref{eq:disp}; they are shown in Fig.~\ref{fig:Figure2}\footnote{In the determination
of these limits, Monte-Carlo simulations were used to evaluate the statistical power of each method.}.
Only those on the linear term
obtained from GRB~090510 with the three methods exceed the Planck energy
scale, the other three GRBs providing bounds an order of magnitude lower.
Including systematic effects at the level of statistical errors (i.e.
introducing a $100$\% error increase into
likelihood calculations) does not change any of the presented conclusions. The
limits on the quadratic term are of the order $10^{11}$~GeV at best, far from the
Planck scale. The limitation of time-of-flight studies with astrophysical
sources arises from the restricted range in $\gamma$-ray energies due to the
opacity of the extragalactic medium which affects their propagation through
space (see Section~\ref{PROBES:propagation})\footnote{Various models describing the EBL which causes energy loss in the propagation of the photons
have been proposed and compared to measurements on high redshift
extragalactic sources~\cite{Franceschini,Kneiske}.}.
Additional studies taking into account a possible dependence of the speed of light on photon helicity leading to much
more stringent limits on LIV are presently underway. \\

In addition to the studies with GRBs, TeV observations of bright AGN flares
provided the opportunity to make LIV studies in complementary energy and redshift
ranges. AGN observed at TeV energies are less variable than GRBs and have only moderate redshifts,
but they provide an extended range of photon energies and usually data samples with higher statistics.
In particular, the TeV range is of great importance for constraining the quadratic term in the dispersion
relation \ref{eq:disp}.

\begin{figure}[!t]
  \centering
  \includegraphics[width=7.5cm]{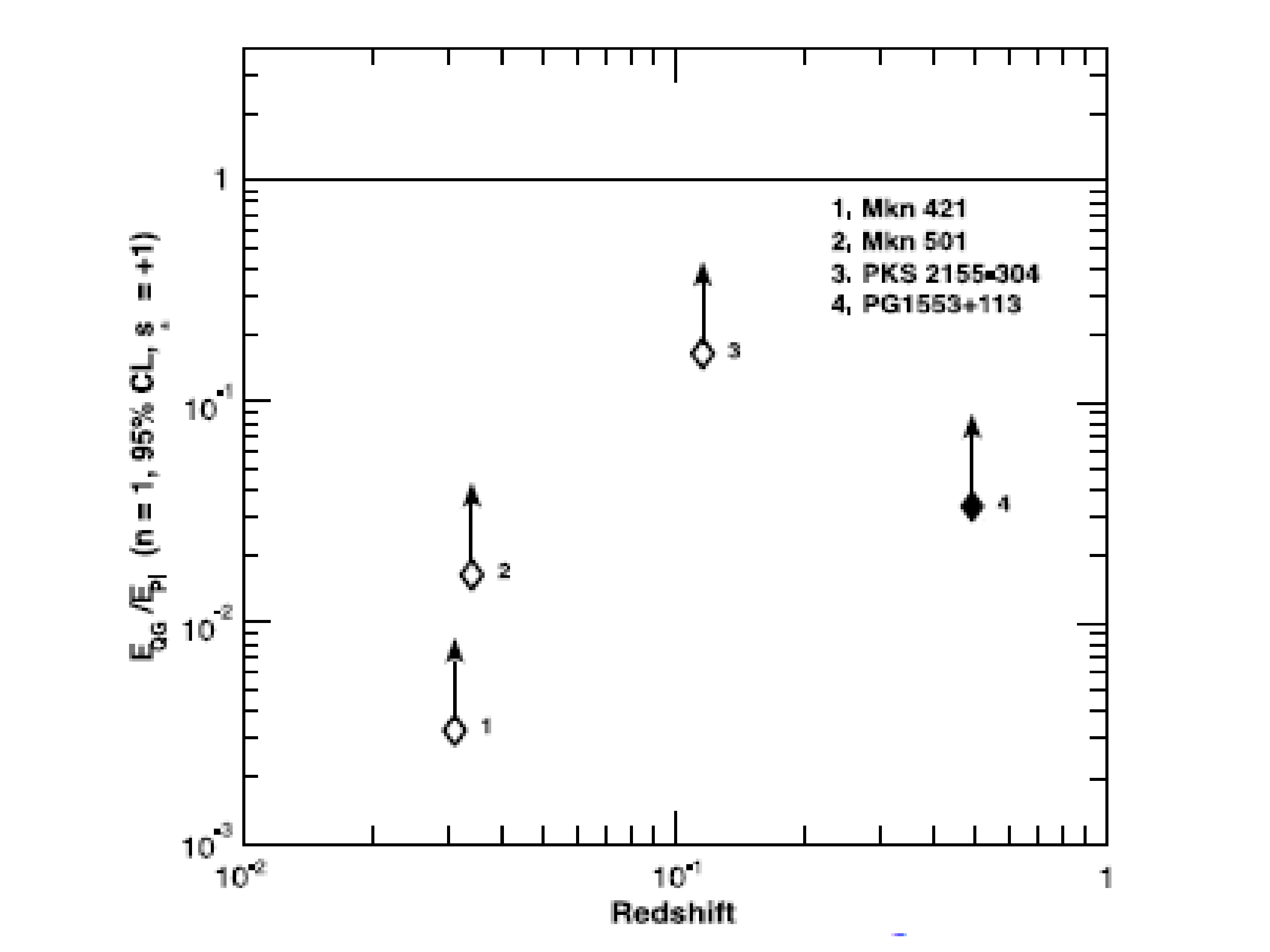}
  \includegraphics[width=7.5cm]{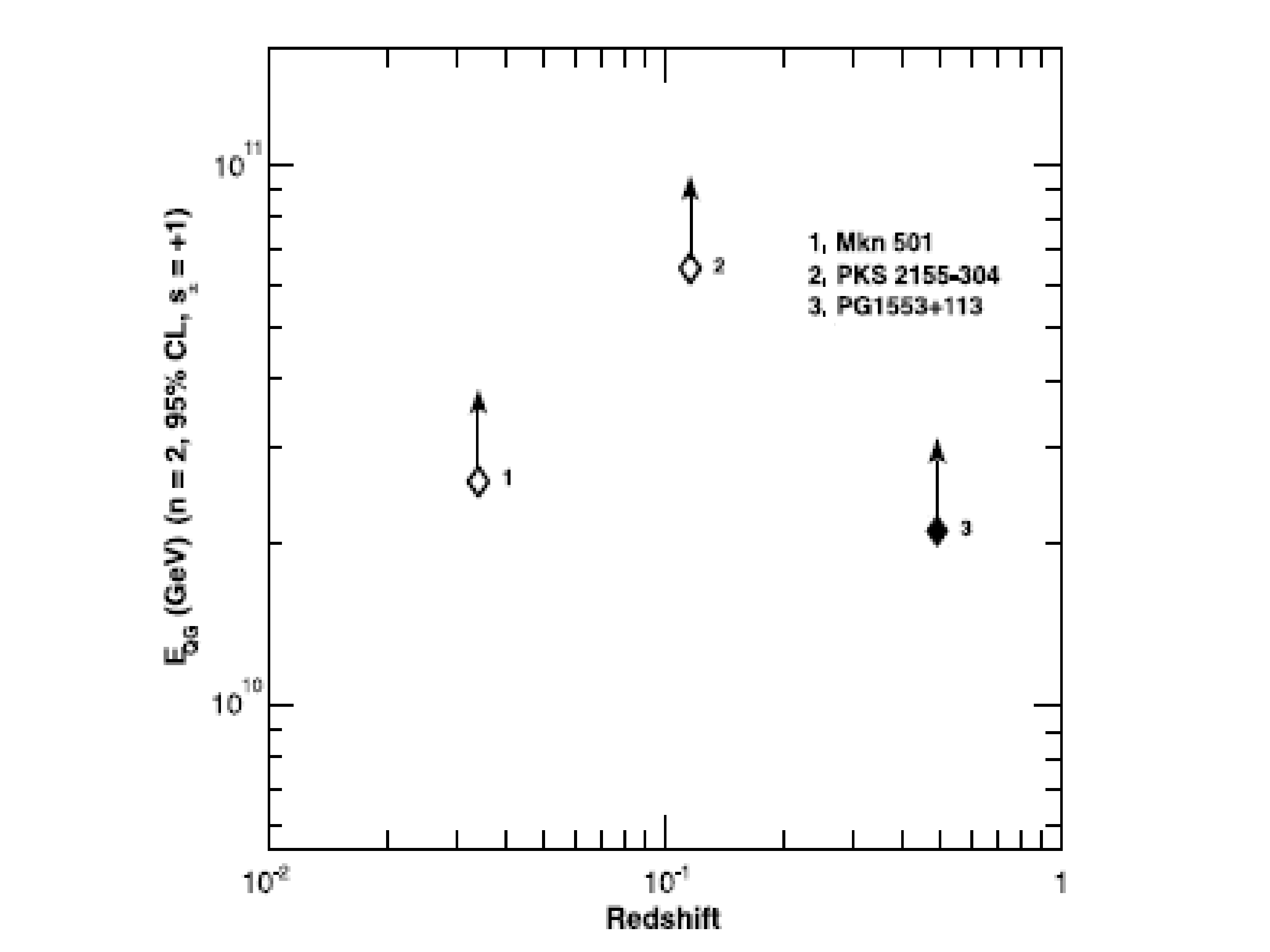}
  \caption{$95$\% CLs on QG energy scale with three AGN data from Whipple, MAGIC and H.E.S.S. (left n = 1, right n = 2). The most performant result was provided with giant flare of PKS 2155-304 in 2006 with thousands of photons were recorded by H.E.S.S. (from~\cite{Abramowski}). The PG 1553+113 results are in press~\cite{Couturier}}.
  \label{fig:Figure3}
\end{figure}

Historically, the first attempt to constrain LIV parameters with VHE photons
from active galaxies was performed by the Whipple Observatory with data from
a Mkn~421 flare in 1996~\cite{Biller}. The obtained constraint on the QG energy
scale was limited by the low statistics of the data sample and the small redshift
($z=0.03$) of Mkn~421 leading to  $E_{QG} > 10^{16}$ GeV, at the level
of those obtained with GRBs detected by the hard X-ray missions at that time, far below $E_{Pl}$.
These studies were followed by further AGN flare analyses by MAGIC and H.E.S.S.
Cherenkov telescopes with an increased potential due to their improved performance.
With the use of
sophisticated statistical methods to extract robust limits on the QG scale,
constraints on LIV effects were found to be competitive with respect to those
obtained from GRBs. The analysis of a flare of Mkn~501~\cite{MAGIC,Martinez} by MAGIC
and that of the exceptional flare of
PKS 2155-304~in 2006 \cite{Aharonian,Abramowski} by H.E.S.S. lead to bounds on the
QG energy scale only an order of magnitude
below $E_{Pl}$, as shown in Fig.~\ref{fig:Figure3}.
Nevertheless, these results are of great interest for the overall physics
picture as they complete the redshift range not covered by the GRB studies.
Moreover, in the future, additional progress could be obtained by
combining GRB and AGN results.\\

The first results of interest on LIV effects obtained from pulsars came with
the discovery of the VHE periodic emission from the Crab pulsar up to
few hundreds of GeV by VERITAS in 2013. Pulsars offer the advantage of a permanent emission
with extremely sharp time variations of the order of
tens of milliseconds. LIV studies with pulsars are submitted to different systematic
uncertainties as compared to GRBs or AGN; in particular,
the slow-down processes of
their emission period in time are well measured and under control. The present
limit obtained from the observation of the Crab pulsar above 100 GeV by VERITAS
is of  $E_{QG} > 0.02 E_{Pl}$ for the linear dispersion
term~\cite{Zitzer}.

%

\begin{table}
\caption{Summary of the results from searches for LIV effects obtained till 2014 with GRBs and AGN. The best limits on the linear dispersion term come from four Fermi GRBs and from the flare of PKS 2155-304 with H.E.S.S. for the linear dispersion term. The best limits on the quadratic term are of the same order of magnitude for GRBs and AGN. Only studies of the four Fermi GRBs and of the flare of PKS 2155-304 take systematic effects into account.}
\begin{center}
\begin{tabular}{|l|c|c|c|}
\hline
Source & Experiment &  $E_{QG}$ (GeV) linear term  & $E_{QG}$ (GeV) quadratic term  \\
\hline
GRB 021206    & RHESSI             &    1.5$\times10^{17}$ GeV  &  \\
9 GRBs           & BATSE + OSSE   &    0.7$\times10^{16}$ GeV  &  \\
15 GRBs         & HETE-2              &    0.4$\times10^{16}$ GeV  &  \\
17 GRBs         & INTEGRAL          &    0.4$\times10^{11}$ GeV  &  \\
35 GRBs         & BATSE + HETE-2 + SWIFT &    1.4$\times10^{16}$ GeV  &  \\
GRB 090510   & GBM + Fermi-LAT &    1.2$\times10^{19}$ GeV  &  0.5$\times10^{11}$ GeV  \\
4 GRBs           & Fermi-LAT &    2.5$\times10^{19}$ GeV  &  0.4$\times10^{11}$ GeV  \\
\hline
Mkn 501         & Whipple             &    0.6$\times10^{16}$ GeV  &  \\
Mkn 501         & MAGIC               &    0.2$\times10^{18}$ GeV  &  0.3$\times10^{11}$ GeV  \\
PKS 2155-304 & H.E.S.S.              &   2.1$\times10^{18}$ GeV  &  0.5$\times10^{11}$ GeV  \\
\hline
\end{tabular}
\end{center}
\label{tab:Table3}
\end{table}

In summary, the results on searches for LIV effects obtained up to 2014 with time-of-flight studies with 
GRBs and AGN are presented in Table~\ref{tab:Table3}, showing the substantial progress
achieved in the last 10 years. The best limits on the linear dispersion term are
those obtained from four Fermi GRBs and from the flare of PKS 2155-304 with H.E.S.S..
The best limits on the quadratic term are of the same order
of magnitude for GRBs and AGN, much below $E_{Pl}$. Only the studies of the four Fermi GRBs
and of the flare of PKS 2155-304 take systematic effects into account.

On the other hand, while differential time-of-flight measurements with HE and VHE photons provide the most
generic and model-independent insight into LIV effects, another approach
based on TeV blazar energy spectra has been proposed
\cite{kifune,piran}. The $\gamma$-ray absorption process by the EBL ($\gamma + \gamma \rightarrow e^+ + e^-$)
could be affected by LIV modifying the energy threshold of the reaction.
However, population studies depending on source redshifts are also needed
in order to disentangle genuine LIV effects from possible
source-induced spectral modifications.
More recently, as mentioned in section \ref{ebl-constraint}, modifications of the pair-production thresholds have been investigated
in order to extract LIV effects from the spectral shapes of
a substantial set of AGN. In this study \cite{biteauwilliams2015}, the
lower limit on the QG scale, found to be around
0.6 Planck scale, can be considered as competitive with those
obtained from time-of-flight measurements on GRBs and AGN until now. However, in this type of analysis combining
measurements from different experiments, the detailed instrumental response must be taken into account since
systematic effects play a crucial role in the determination of the limits on LIV parameters\footnote{It should be
noticed here, that blazar
energy spectra are also basic ingredients of ALP studies, thus implying a
close connection between the LIV and the ALP induced modifications, see the discussion in the end of this article.}.

\subsubsection{Discussion of results on LIV and future prospects}
\label{sec:discussion}

One limiting, not yet well studied problem in the interpretation of the results
concerns the impact of the emission processes at the source, which depends on the type of object.
As an example, it is known that GRBs suffer from intrinsic
lags~\cite{Fenimore,Norris} as observed already by BATSE: at higher
energies, emission peaks at shorter times and photons arrive earlier. 
Correlations were observed between time-lags and peak luminosity in the
keV-MeV range. First tests with Fermi GRBs in GeV energy range including spectral variations have shown no major change
in the LIV results and values of the QG energy scale $E_{QG}$. These effects, related to the
underlying mechanism of the GRB emission need further studies with more sources in the analysis.

The search for the LIV-induced effects in the
light-curves and spectra of GRBs, AGN and pulsars assuming
a deterministic dispersion relation in vacuum made big progress in
the last ten years. The study on GRB 090510 provided the most stringent and
experimentally solid bound above Planck scale for the linear
term, but not an ultimate one, as a unique event. Other results with GRBs, AGN and
pulsars, point at limits an order of magnitude below the Planck
scale. To make our conclusion robust on the physics side and discuss the validity
of proposed theoretical models, more observations of exceptional events as
the prompt GRB 090510 or giant flares of high redshift AGN as of the one
observed on PKS 2155-304 in 2006 are needed. Also, a combination of already
existing results taking into account the redshift of each source would be a
step forward in this domain. The stacking procedure would provide a most robust
result on LIV effects averaged over systematic uncertainties and almost free from
the internal time-delays of each source.

\subsection{Phenomenology of photon-ALP mixing}
\label{PROBES:sub:pheno_ALPS}
While in the previous section, effects of LIV on the arrival time of photons are considered,
we now turn to the possible modification of the optical depth for $\gamma$-ray sources. Such an effect
can be induced by photons mixing with new particles which are well-motivated phenomenologically
as well as theoretically. The axion field was introduced in the framework of an extension of the Standard Model,
in order to solve a problem related to CP conservation in strong interactions \cite{pecceiquinn1977}.
This conservation is puzzling since, in the Quantum Chromodynamics (QCD) Lagrangian, a CP-violating term is expected, of the form:
\begin{eqnarray}
\mathcal{L}_\mathrm{strong,CP}&=& \frac{\alpha_s}{8\pi} \, \bar \theta \, G_{a\mu\nu}\tilde G_a^{\mu\nu}
\end{eqnarray}
in which $\alpha_s$ is the strong coupling constant, $G$ represents the QCD field tensors and $\tilde G$
its dual, and $\bar{\theta}$ is an unpredicted CP violating phase $-\pi \le \bar \theta \le \pi$. A priori, a value
$\bar \theta = \mathcal{O}(1)$ could be expected. However, the experimental fact that the electric
dipole moment of the neutron is extremely small ($d_n< 2.9\times 10^{-28}~e\cdot \mathrm{m}$ at
90\% confidence limit \cite{baker2006}) translates into the constraint $|\bar \theta| < 10^{-10}$. The
fine-tuning of $\bar \theta$ was elegantly solved by Peccei and Quinn by adding a new spontaneously broken global symmetry
$U(1)_\mathrm{pq}$ \cite{pecceiquinn1977} with the resulting
so-called \textit{axion} field $A$, with decay constant $f_A$, cancelling out dynamically the CP-violating term according to the modified Lagrangian:
\begin{eqnarray}
\mathcal{L}_\mathrm{strong, CP}&=& \frac{\alpha_s}{8\pi}
\left(\bar \theta -
\frac{A}{f_A}\right)
G_{a\mu\nu}\tilde G_a^{\mu\nu},
\end{eqnarray}
$A$ reaches its minimum value $\bar \theta f_A$
through non-perturbative effects in QCD.

The oscillation of the field around its minimum
gives rise to the axion as a pseudo Nambu-Goldstone boson
whose mass is fixed to
$m_A\approx 0.6~\mathrm{meV} (f_A/10^{10}~\mathrm{GeV})^{-1}$,
with some model dependence on how the additional $U(1)$ symmetry is broken.
For a review on the QCD axion, see e.g.
\cite{Ringwald2014}\footnote{For a simple analogy
of the Peccei Quinn mechanism with a snooker table see \cite{Sikivie95}.}. \\
As a pseudo-scalar boson, the axion $a$ can be coupled to photons through the Primakoff effect\footnote{This effect
is well-known in the production of pseudo-scalar mesons such as $\pi^0$s.}
($\gamma + $ (possibly virtual) $\gamma \rightarrow a$).
The mixing of axions (pseudo-scalar boson) with photons (vector bosons) requires
an external field, the axion-two photon interaction being described by the Lagrangian:
\begin{eqnarray}
  \mathcal{L}_{A\gamma\gamma} &=& -\frac{G_{A\gamma\gamma}}{4} A F_{\mu\nu}
\tilde{F}^{\mu\nu} = G_{A\gamma\gamma} \mathbf{E}\cdot\mathbf{B},
\end{eqnarray}
where $F_{\mu\nu}$ is the electromagnetic field tensor and $\tilde{F}^{\mu\nu}$
its dual. The coupling constant $G_{A\gamma\gamma}\propto 1/f_A\propto m_A$ is
proportional to the mass of the axion. Intensive searches using
both laboratory experiments and astrophysical observations have lead to
constraints on the photon-axion coupling and therefore on its mass to be below
few $10~\mathrm{meV}$ (for an overview see \cite{Ringwald2014}).

\begin{figure}[t!]
\centering
 \includegraphics[width=14 cm]{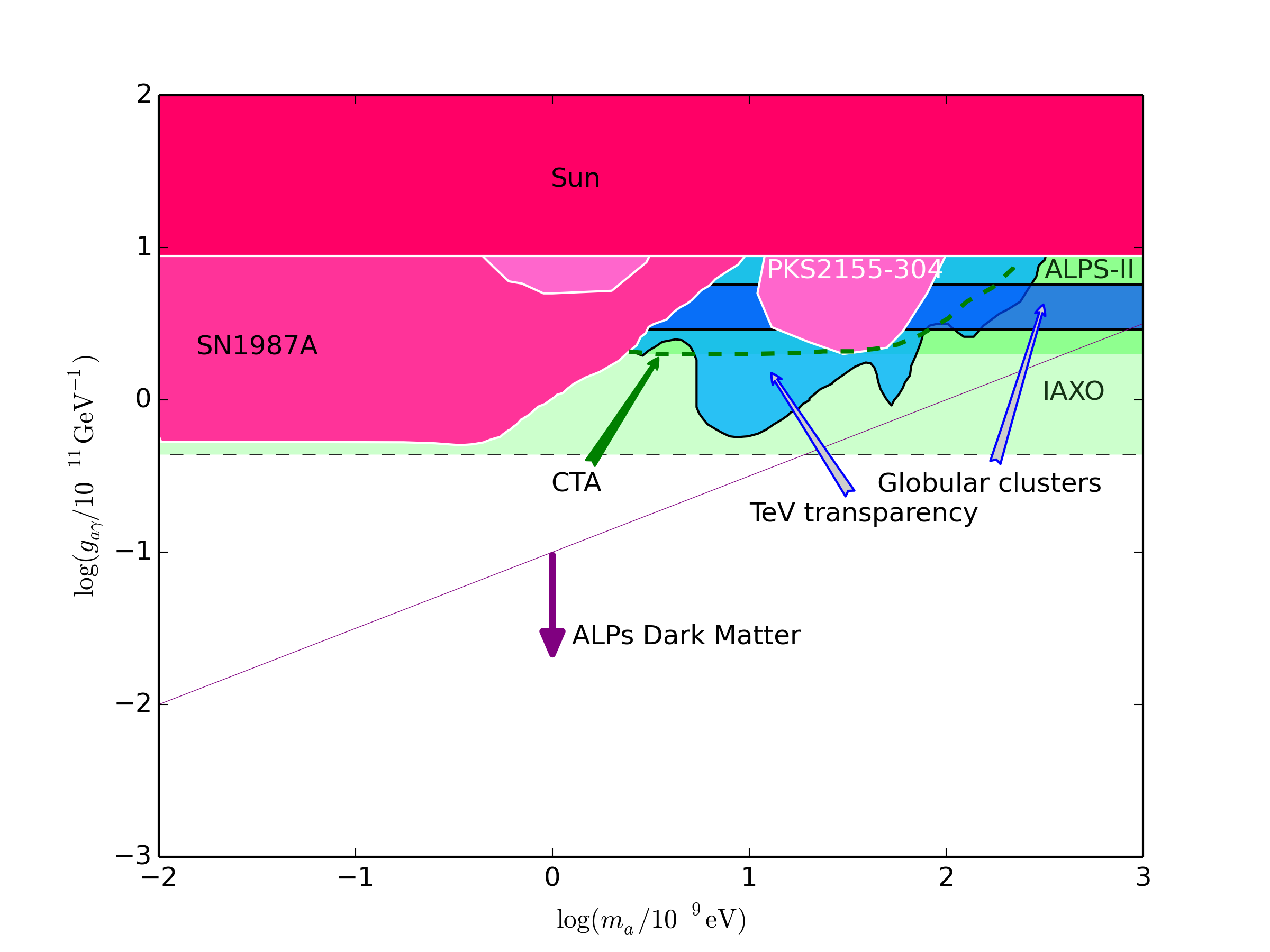}
 \caption{ \label{fig:alps_plane}
ALPs parameter space: coupling ($g_{a\gamma}$) versus mass ($m_a$). The plot is centered
on the region where $\gamma$-ray observations are sensitive. The diagonal
line marks the approximate upper bound of the region of parameter space which
would explain the non-baryonic dark matter present in the universe \cite{Arias2012}.
Experimental
bounds from observations of the sun (CAST experiment \cite{andriamonje2007}),
of the active galactic nucleus PKS~2155-304 (H.E.S.S. experiment
\cite{abramowski2013b}),
and the non-observation of $\gamma$-rays from SN~1987 (Solar maximum explorer
\cite{payez2014}) are indicated by white boundaries, the pink regions being excluded by
these observations. The black lines indicate boundaries of the parameter space
which are favored by observations. The enclosed blue regions would explain
the anomalous transparency of the universe to TeV $\gamma$-rays \cite{meyer2013b}
and the ratio of horizontal branch stars to red giant stars in globular clusters
\cite{ayala2014}.
One should note the considerable overlap in parameter space
of these two independent measurements.
The dashed lines indicate the sensitivity of future experiments including the ``light-shining through the wall'' experiment ALPS-II \cite{baehre2013},
the planned helioscope IAXO \cite{armengaud2014}, and
the future ground-based $\gamma$-ray telescope system (CTA) \cite{meyer2014}.}
\end{figure}

In a more generic fashion, additional fundamental pseudo-scalars are motivated
in specific string compactification models \cite{Cicoli2012}, where,
besides the QCD axions, further fundamental pseudo-scalars could exist
with a logarithmic mass hierarchy
(for a review, see e.g., \cite{jaeckelringwald2010}).
These so-called \textit{axion-like} particles (ALPs) would share similar
couplings to photons via
$\mathcal{L}_\mathrm{a\gamma\gamma}=-1/4\,g_{a\gamma}\,a~\mathbf{E}\cdot\mathbf{B}$, with
$a$ being the ALP-field,  $g_{a\gamma}$ its coupling, $\mathbf{E}$ the  {electric} field, e.g., of a propagating
electro-magnetic wave, and $\mathbf{B}$, e.g., an external magnetic field.
\\
For ALPs, there
is generally no a priori constraint favoring particular regions in the
parameter space
of mass $m_a$ and coupling $g_{a\gamma}$, unless more specific assumptions
on theoretical grounds are included.
The above mentioned string compactification model would favor
$\mathcal{O}(100)$ different ALPs at logarithmically spaced mass values with couplings
between $\approx 10^{-13}~\mathrm{GeV}^{-1}$
and a few $10^{-11}~\mathrm{GeV}^{-1}$ \cite{Cicoli2012}. \\
In Fig.~\ref{fig:alps_plane}, a region of the ALPs-parameter space is shown
which is relevant for $\gamma$-ray observations.
In case of a non-thermal cosmological production of ALPs through the vacuum alignment mechanism \cite{abbott1983},
ALPs represented below the diagonal magenta-colored line labelled \textit{ALP DM} in Fig.~\ref{fig:alps_plane} would be candidates for
non baryonic cold dark matter\footnote{One should note that the previously described
axion which solves the CP-problem would be represented outside of the boundaries of the plot. Similarly,
searches for ALPs dark matter using micro-wave cavities (so-called haloscopes) start at larger mass values.}.
The most important and strongest constraints are included in the plot as white lines.
Values of coupling represented above the white line (pink shades) have been excluded through the following studies:
\begin{itemize}
  \item searches for ALPs escaping from the hot core of the sun and re-converting
into X-ray photons in a transverse magnetic field of a helioscope (e.g.,
the CAST experiment \cite{andriamonje2007})\footnote{A
similar constraining bound not shown in Fig.~\ref{fig:alps_plane} has been derived from the
existence of the blue-loop phase in the evolution of massive stars
\cite{Friedland2012}};
  \item non-detection of
$\gamma$-rays from the collapsing proto-neutron star in SN1987 \cite{payez2014};
  \item searches for irregularities in the $\gamma$-ray spectrum of the AGN PKS2155-304
\cite{abramowski2013b}, see also Section~\ref{sub:noise}.
\end{itemize}
The sensitivity of future searches is indicated by light/green shaded regions
for the next-generation ``light-shining through a wall'' experiment ALPS-II \cite{baehre2013}
and the planned helioscope IAXO \cite{armengaud2014}.  The existing hints
for the existence of ALPs are marked as black lines with the blue shaded
areas favored by the observational data, e.g., on an anomalous transparency with respect to $\gamma$-rays
\cite{horns2012,meyer2013b} and on the live-time of horizontal branch stars
in globular clusters \cite{ayala2014}.\\

If nature is so generous to provide these additional
fields as the remnants of physics at a higher energy scale, $\gamma$-ray observations
would be sensitive to ALPs at a mass scale of $m_a<10^{-6}$~eV and couplings as low as $g_{a\gamma}=\mathcal{O}(10^{-11}~\mathrm{GeV}^{-1})$. This mass range is mainly fixed through the external magnetic fields which affect the mixing of photons and ALPs.

The theoretical problem of  $\gamma$-rays propagating in the presence of an ALP
and external magnetic field
has been considered by a number of authors since \cite{raffelt88};
a recent and fairly complete description of the
problem and its solution including absorption effects has been given in
\cite{mirizzi2009,deangelis2011}.
The photon state described by the polarization vector $\left\{ \epsilon_1,\epsilon_2 \right\}$
introduced in Section~\ref{PROBES:propagation}
mixes in the presence of an external magnetic field  with the ALP state~$a$. The resulting evolution of a
photon/ALP-state $\left\{\epsilon_1,\epsilon_2,a\right\}$
propagating in the presence of an external magnetic field $\mathbf{B}$ is described by
\cite{czaki2003,mirizzi2009}:
\begin{eqnarray}
-i\frac{\partial}{\partial x_3}
\left( \begin{array}{c} \epsilon_1 \\ \epsilon_2 \\ a\end{array}\right)&=&
\left( \begin{array}{ccc}
\Delta_{11}-i\frac{\lambda^{-1}_{\gamma\gamma}}{2} &  \Delta_{12} & \Delta_{a\gamma}\cos\varphi \\
\Delta_{21}          &  \Delta_{22}-i\frac{\lambda^{-1}_{\gamma\gamma}}{2} & \Delta_{a\gamma}\sin\varphi \\
\Delta_{a\gamma} \cos\varphi     & \Delta_{a\gamma}\sin\varphi         & \Delta_{a}  \\
\end{array} \right)
\left( \begin{array}{c} \epsilon_1 \\ \epsilon_2 \\ a\end{array}\right),
\end{eqnarray}
where $\Delta_{11}$, $\Delta_{12}=\Delta_{21}$,  $\Delta_{22}$, and $\varphi$,
already defined in Sect.~\ref{PROBES:propagation},
{are related to} the effect of the {magnetized} background plasma,
and to the polarization of the vacuum on the propagating beam.
Effects of the Faraday rotation can be neglected at the
high energies considered here. The mixing of photons with ALPs in the beam is mediated
by the term:
\begin{eqnarray}
\Delta_{a\gamma}&=&\frac{1}{2} g_{a\gamma} |\mathbf{B}(1-\mathbf{e}_3)|\approx 1.52\times 10^{-2}\left(\frac{g_{a\gamma}}{10^{-11}~\mathrm{GeV^{-1}}}\right) \left(\frac{B_\perp}{10^{-13}~\mathrm{T}}\right)~\mathrm{Mpc}^{-1}.
\end{eqnarray}

For a homogeneous medium (constant field $\mathrm{B}$, constant plasma density) and in the absence
of absorption, the probability $p_{\gamma\rightarrow a}$ of a photon state to convert into an ALP-state after propagating
a distance $L$ is given by
\begin{eqnarray}
  p_{\gamma\rightarrow a} &=& \sin^2(2\vartheta)~\sin^2\left(\frac{L\Delta_\mathrm{osc}}{2}\right),
\end{eqnarray}
with  the oscillation wave number $\Delta_\mathrm{osc}$ and mixing angle $\vartheta$ given by:
\begin{eqnarray}
   \Delta_\mathrm{osc}&=& 2\Delta_{a\gamma} \sqrt{1+\left(\frac{E_c}{E}\right)^2}
\end{eqnarray}
and
\begin{eqnarray}
\sin(2\vartheta)&=&\frac{2\Delta_{a\gamma}}{\Delta_\mathrm{osc}} = \frac{1}{\sqrt{1+\left(\frac{E_c}{E}\right)^2}}
\end{eqnarray}
respectively. At energies exceeding the critical energy $E_c$
\begin{eqnarray}
 E_c &=& \frac{|m_a^2 - \omega_\mathrm{pl}^2|}{4 \Delta_{a\gamma}}
\approx 2.5\times 10^{-1}~\mathrm{GeV}\frac{|m_a^2 - \omega_\mathrm{pl}^2|}{(10^{-9}~\mathrm{eV})^2}
\left(\frac{10^{-13}~\mathrm{T}}{B_\perp}\right) \left(\frac{10^{-11}~\mathrm{GeV}^{-1}}{g_{a\gamma}}\right)
\end{eqnarray}
the mixing angle $\vartheta=\pi/4$ leads to maximum mixing, independent of the
energy of the photon.
The critical energy separates the
 strong-mixing regime ($E\gg E_c$) from the no-mixing regime ($E\ll E_c$) with a
transition region ($E=\mathcal{O}(E_c)$).
In the strong mixing regime $\Delta_\mathrm{osc}\approx 2\Delta_{a\gamma}$ and
the corresponding oscillation
length $\Delta_\mathrm{osc}^{-1} \approx (2\Delta_{a\gamma})^{-1}=\mathcal{O}(100)~\mathrm{Mpc}$ for
magnetic fields of less than $10^{-13}~\mathrm{T}$ typical for the intergalactic medium. The propagation in
an inhomogeneous medium with varying strength and orientation of the magnetic field can be treated with
a cell-like approach, where the photon-ALP beam is propagated over cells of size similar to that expected for the coherence
length of the magnetic field $\mathcal{O}(\mathrm{Mpc})$ \cite{mirizzi2009}.
Alternatively, the expected turbulent spectrum of  magnetic
field strength can be considered directly via Fourier transformations as suggested in \cite{meyer2014b} with
results similar to the cell-like approach.

In the past decade, a number of approaches have been suggested and used to
search for signatures of PAM through observations of $\gamma$-ray sources, which
are detailed in the following sections.
\subsubsection{Anomalous transparency for $\gamma$-ray photons}
\label{anomalous}
Provided that the conversion probability $p_{\gamma\rightarrow a}\times p_{a\rightarrow\gamma}$
exceeds the probability for absorption,
$\gamma$-rays can effectively mix with ALPs before
being attenuated due to pair-production and further re-convert into photons closer to the observer.
In the equilibrium case of strong mixing, one third of the intensity of the photon beam
will be propagating effectively as an ALP beam. The initial conversion of
$\gamma$-rays can take place in the vicinity of the emitting region, e.g.,
in AGN jets \cite{sanchez2009,tavecchio2012,tavecchio2014},
in the magnetized intergalactic plasma of galaxy clusters hosting AGN \cite{horns2012}, or
even in the magnetic field of the intergalactic medium \cite{mirizzi2009}.
While the photonic part of the beam is on average attenuated, the ALP beam
does not suffer from any losses.
When the beam propagates in the magnetic field of our host galaxy, the conversion {$p_{a\rightarrow \gamma}$} will lead to the appearance of
photons in excess of the attenuated beam. \\
This mechanism leads to a modification of the optical depth in the
part of the observed energy spectra submitted to absorption. The energy at which the
optical depth is sufficiently large depends on the distance of the source as well as on
the structure of the intervening magnetic field. Given the unknown strength of the
transverse field, it is not predictable and leads to a stochastic process of conversion/re-conversion,
first pointed out in \cite{mirizzi2009}. \\
The effect has been
claimed to be present consistently in various analyses of $\gamma$-ray spectra
observed with atmospheric Cherenkov telescopes \cite{deangelis2009,deangelis2011,horns2012b,galanti2015}.
Subsequent discoveries of a number of additional sources
strengthen these indications \cite{rubtsov2014,meyer2013}. Furthermore, the observation of
$\gamma$-ray-like extensive air showers at energies $>2\times 10^{17}$ GeV has been suggested to be
consistent with PAM \cite{fomin2013}.
{As discussed previously,  the spectral energy density of the EBL
has been extracted in the range of $0.2-100~\mu$m using
$\gamma$-ray spectra of 38 AGN \cite{biteauwilliams2015}. Even
though the authors claim that they find no direct indications for a
pair-production anomaly in the extended data-set, the resulting
EBL has however two surprising features: a large intensity in
the lowest wavelength interval (see Fig.~\ref{ebl}), which is not expected from
any EBL model and tension with the lower bounds from galaxy counts in the near to mid infrared.
Furthermore, their model dependent estimate of the Hubble constant deviates by $2.6~\sigma$ (statistical uncertainties only)  from
the value inferred using e.g. the Planck data \cite{planck2014}. 
Even though it is too early to draw firm conclusions, the new study demonstrates
that the EBL inferred from $\gamma$-ray observations only, may require additional
explanations and could point towards an
anomalous transparency of the Universe to VHE $\gamma$-rays, either through PAM or
LIV.} Such indications are
consistent with the PAM scenario studied in
\cite{meyer2013b} where
the photon-ALP conversion takes place predominantly in the
magnetic field of the source environment, while the back-conversion
takes place in the Galactic magnetic field.
The favored regime of coupling and mass, i.e. the
region marked \textit{TeV transparency} in Fig.~\ref{fig:alps_plane},
is not ruled out by current laboratory experiments.\\

The re-conversion of the un-attenuated ALP beam in the Galaxy
depends on the strength of the transverse Galactic magnetic field. Even though
the large scale structure of the Galactic magnetic field has been studied
through polarization data of background radio-galaxies (see e.g., \cite{jansson}),
the constraints on the strength and orientation of the field outside
of the Galactic plane are poorly known. Independently of the particular assumption
on the magnetic field structure, it is clear that the re-conversion probability
depends on the line-of-sight towards the $\gamma$-ray source. This in turn
is a unique signature for the interpretation of $\gamma$-ray data with
PAM \cite{simet2008}. However, the sensitivity of current instruments is
not sufficient to detect the directional dependence.
Future observatories such as Cherenkov Telescope Array (CTA) \cite{FUTURE}, even with
a rather small data-set of four $\gamma$-ray sources, will be sensitive to
couplings as low as $g_{a\gamma}>2\times 10^{-11}~\mathrm{GeV}^{-1}$ and $m_a<100~\mathrm{neV}$ \cite{meyer2014}.

\subsubsection{Excess noise in $\gamma$-ray spectra}
\label{sub:noise}
The studies discussed so far have been considering the regime of strong mixing where
the transparency for pair-production is modified ({$E\gg E_\mathrm{c}$}).
At smaller energies, the mixing is less efficient and the
propagation in a turbulent/randomized magnetic field leads to additional fluctuations in the observed energy spectra. This effect was first studied
in the context of optical spectra from distant quasars \cite{ostman2004}.
For $\gamma$-ray spectra, this approach has been used by the H.E.S.S. collaboration
to exclude ALP parameters (i.e. in Fig.~\ref{fig:alps_plane} the regions marked
\textit{PKS~2155-304}) from the well-measured spectrum of
a low redshift AGN ($z=0.116$) \cite{abramowski2013b}.
This method is only sensitive in a narrow mass range, because the irregularities
are produced for photons of energies similar to the critical energy. Furthermore,
the method is more sensitive to the assumed magnetic field; therefore, the two pink exclusion
patches shown in Fig.~\ref{fig:alps_plane} are derived for two different types of magnetic fields.

\subsubsection{Disappearance of Galactic $\gamma$-ray emission}
While the conversion/re-conversion leads to the appearance of photons in the
optically thick part of energy spectra, the mixing can also lead to an energy-dependent
suppression because of photons converting to undetected ALPs.
The photon-ALP conversion
depends strongly on the line-of-sight orientation with respect to
the large scale magnetic field in the Galaxy. This is specifically
true for photon beams propagating across
different spiral arms. While the propagation in the turbulent magnetic field of
galaxy clusters or in the intergalactic medium is not predictable, only
the regular
component of the Galactic magnetic field is of importance here since, for its
turbulent part, the conversion probability is too small to matter.
Therefore, the effect can be predicted and an actual set of parameters (mass and coupling)
can be determined quite accurately. Initial analyses of HE $\gamma$-ray spectra from Galactic
pulsars are encouraging \cite{horns2014} and can be extended to other Galactic sources.

\subsubsection{Discussion and outlook}
The $\gamma$-ray observations of sources at very large distances have revealed a puzzling effect where
energy spectra tend to be less affected by absorption than predicted in the
framework of standard photon propagation. 
New experiments with more data and better instrumentations, (e.g. Cherenkov
Telescope Array (CTA)) are needed. The notable differences between the two scenarios (ALPs vs. cascades)
would be the following:
\begin{itemize}
\item Observation of a variability of distant $\gamma$-ray sources that is not predicted if
a substantial part of the emission is produced in cascades. The latter tend to smear out the
variability over time-scales of years \cite{prosekin2012}.
\item Angular dependence of the re-conversion effect. In the ALP-scenario,
the re-conversion of ALPs in the propagating beam is efficiently taking place in the
Galactic  magnetic field. Even though the precise structure of the field is not well known,
differences in the spectra due to PAM effects will be observable using a large sample of AGN at various distances
\cite{wouters2014}.
\item Conversion in the Galactic magnetic field. With sufficiently precise spectra from various
Galactic sources, the expected coupling and mass from the anomalous transparency will lead
to modulations (photon disappearance at some energies) for Galactic $\gamma$-ray sources where the line-of-sight
happens to cross spiral arms. In other sources, this effect should be absent \cite{horns2014}.
\end{itemize}
Finally, other independent observations of e.g., horizontal branch stars in globular clusters
\cite{ayala2014} or of linear polarization of optical emission from magnetic white dwarfs
\cite{Gill2011} have the potential to probe the same parameter space of photon-ALP coupling
as favored from the anomalous transparency observations. A combined analysis of the
available data is pointing towards a consistent picture of new physics at the intermediate energy
scale \cite{dias2014}.

\section{Summary and general discussion}
\label{PROBES:summary}

The analyses of a large sample of astrophysical sources, as described in
this paper, has brought a new insight into {the
amount of optical/infrared background light as well as} in the area of fundamental physics
by testing otherwise well established symmetries and possible implications of their breaking
at different energy scales. These searches do not require new or specific
astrophysical observations, still leading to results of great
importance. \\

{The results on the extragalactic background light
inferred from absorption features in $\gamma$-ray spectra have helped
clarify the amount of optical/infrared light present in the universe.
This in turn can be used to constrain the star-formation history and
contributions from new phenomena injecting additional light in the early
universe.}\\

At present, the limits on the energy scale of Quantum Gravity, obtained with a good control of systematics,
are close to the Planck scale
in case the speed of light varies linearly with photon energies. On the other hand, the quadratic term in the dispersion
relations is poorly constrained and needs not only further
observations but also a better sensitivity of the experiments in the highest energy
part of spectra taking EBL attenuation into account.
Veryfing the constancy of the speed of
light in vacuum is a crucial test of a cornerstone assumption of the Einstein's theory.
The new potential will come from the CTA project \cite{FUTURE}, where
population studies will provide a gain in sensitivity of an order of
magnitude in the energy range of interest, leading to limits on the linear term
easily reaching the Planck scale for many sources and, for the quadratic term~\cite{Bolmont2},
to bounds two orders of magnitude higher than the present ones.\\

The spectroscopy of distant AGN spectra at energies well beyond the
$\gamma$-ray horizon, where the optical depth $\tau_{\gamma \gamma}$ is greater than unity, is sensitive to
a possible anomalous transparency of the Universe.
Much progress in this field has been achieved in the past years due to a refined understanding of the
attenuating background photon field, and to an increase in the number of objects where
$\gamma$-ray emission is detected at a level above that expected from the
absorption process \cite{horns2012}. This has been interpreted in the context of photon-ALP mixing
(PAM) \cite{meyer2013b}.
The coupling $g_\mathrm{a\gamma}$ between photons and a
new light ($m_a<10^{-6}~\mathrm{eV}$) pseudoscalar boson required to explain the
observations, points towards new physics at energies in the range $10^{11} - 10^{12}$~GeV,
a scale intermediate between that of electro-weak symmetry breaking and
the Planck scale. Even though
the anomalous transparency could be interpreted in different scenarios as well
(e.g. LIV \cite{fairbairn14} or secondary emission of cascades \cite{essey2010}),
the theoretical understanding gives clear predictions which will help distinguish between different interpretations in the future.
The observation of various independent phenomena related to the
propagation of $\gamma$-rays clearly demonstrates that the sensitivity to
PAM is comparable to or even better than any other experimental observations.
The unique mass and coupling regime probed by $\gamma$-ray observations is
at the same time accessible to the next generation of helioscopes (IAXO
\cite{armengaud2014}) and ``light-shining-through the wall'' experiments
(ALPS-II \cite{baehre2013}). Even the existing CAST experiment after substantial upgrades of the detector system has started 
to probe the regime favored by the anomalous transparency observations \cite{lakic2013}. 

Finally, it should be underlined that those searches for deviations from the Standard Model based on the propagation of energetic photons from astrophysical sources require taking account of various cross-correlated effects when interpreting results.
In particular, a clear connection exists between LIV and PAM effects as presented in this paper.

\section*{Acknowledgements}
The authors are grateful to Steven Fegan for his careful reading of the manuscript and his help to improve the overall presentation. 
We also thank Bernard Degrange and G\'erard Fontaine for interesting discussions on the article content.


\end{document}